\def\year{2020}\relax
\documentclass[letterpaper]{article} % DO NOT CHANGE THIS
\usepackage{aaai20}  % DO NOT CHANGE THIS
\usepackage{times}  % DO NOT CHANGE THIS
\usepackage{helvet} % DO NOT CHANGE THIS
\usepackage{courier}  % DO NOT CHANGE THIS
\usepackage[hyphens]{url}  % DO NOT CHANGE THIS
\usepackage{graphicx} % DO NOT CHANGE THIS
\usepackage{graphicx}  % DO NOT CHANGE THIS
\usepackage{amsmath,amssymb,amsfonts}
\usepackage{algorithmic}
\usepackage{textcomp}
\usepackage{xcolor}

\usepackage{booktabs} % For formal tables

\usepackage{array, multirow}
\usepackage{comment}
\usepackage{subcaption}
\usepackage{gensymb}
\usepackage{epstopdf}
\usepackage{enumitem}
\usepackage{tablefootnote}

\usepackage{url}
\makeatletter
\g@addto@macro{\UrlBreaks}{\UrlOrds}
\makeatother

\newcommand{\tabincell}[2]{\begin{tabular}{@{}#1@{}}#2\end{tabular}}
\graphicspath{{./fig2/}}

\newlength\Origarrayrulewidth

\newcommand{\Cline}[1]{%
	\noalign{\global\setlength\Origarrayrulewidth{\arrayrulewidth}}%
	\noalign{\global\setlength\arrayrulewidth{2pt}}\cline{#1}%
	\noalign{\global\setlength\arrayrulewidth{\Origarrayrulewidth}}%
}

\newcommand\Thickvrulel[1]{%
	\multicolumn{1}{!{\vrule width 2pt}c|}{#1}%
}

\def\BibTeX{{\rm B\kern-.05em{\sc i\kern-.025em b}\kern-.08em
		T\kern-.1667em\lower.7ex\hbox{E}\kern-.125emX}}

\title{Cannot Predict Comment Volume of a News Article before (a few) Users Read It}

\author{
	Lihong He$^*$, Chen Shen$^*$, Arjun Mukherjee$^+$, Slobodan Vucetic$^*$, Eduard Dragut$^*$ \\ 
	$^*$Department of Computer and Information Sciences, Temple University \\
	$^+$Computer Science Department, University of Houston \\
	$^*$\{lihong.he, chen.shen, vucetic, edragut\}@temple.edu \\
	$^+$arjun@cs.uh.edu
}

\begin{document}

\maketitle

\begin{abstract}
Many news outlets allow users to contribute comments on topics about daily world events. News articles are the seeds that spring users' interest to contribute content, i.e., comments. 
An article may attract an apathetic user engagement (several tens of comments) or a spontaneous fervent user engagement (thousands of comments).
In this paper, we study the problem of predicting the total number of user comments a news article will receive.
Our main insight is that the early dynamics of user comments contribute the most to an accurate prediction, while news article specific factors have surprisingly little influence. This appears to be an interesting and understudied phenomenon: collective social behavior at a news outlet shapes user response and may even downplay the content of an article.
We compile and analyze a large number of features, both old and novel from literature. 
The features span a broad spectrum of facets including news article and comment contents, temporal dynamics, sentiment/linguistic features, and user behaviors. 
We show that the early arrival rate of comments is the best indicator of the eventual number of comments. We conduct an in-depth analysis of this feature across several dimensions, such as news outlets and news article categories.
We show that the relationship between the early rate and the final number of comments as well as the prediction accuracy vary considerably across news outlets and news article categories (e.g., politics, sports, or health). 
%Our qualitative analysis shows interesting insights related to such a variability.
\end{abstract}

\section{Introduction}
\label{sec:intro}

Commenting on news is a common form of participation in contemporary news consumption, and it is one of the most common forms of citizen engagement online \cite{EmmerVW11}.
A key indicator of user participation in daily news events is the volume of user comments reacting to a news article \cite{ProchazkaWS18,ZiegeleWQB}. Several works propose methods to predict it  \cite{tsagkiasWD09,balaliAF17}. They use a large number of features, which can be broadly categorized into article content, meta-article (e.g., outlet or category), temporal (e.g., date and time of publication), and semantic (e.g., named entities). They model the prediction problem as a classification problem. For example, they determine if an article will receive a ``high" or ``low" volume of comments. %They report two key findings: (1) that in many cases (outlets), the prediction with those features is no better than a random predictor and (2) that among those features, predictors based on article content features alone are the best performers.
One of their key findings is that \textit{(i) 
%among those features, 
predictors based on article content features alone are the best performers and (ii) one may even achieve high accuracy with such predictors.}
%Previous works from the point of computer science find that the content of a news article and its meta information (e.g., article title, publication time, news outlet, category, topic) are good predictors for the volume of user comments.

The social science community, in particular the Communication community, argues that quality discourse emerges only when many users participate in commenting on a news article and when there is interactivity among users, i.e., users comment/reply to prior comments \cite{kiousis02}. The general questions pursued in this space aim to understand the factors affecting participation and interactivity in the comment section of an article \cite{altheideS12,weber14}.
Some studies (via face-to-face interviews) show that factors from previously posted user comments affect the involvement of new users and ultimately increase users' willingness to engage in online news discussions \cite{mishneG06,ZiegeleQ13}. \textit{They conclude that a large fraction of the comments an article receives-- up to 50\%-- do not respond to the journalistic value of a news article, but rather to a previously posted user comment} \cite{singer09,RuizDMDMM11}. 
%They argue that the discussion value of online news articles (or comment volume) is given by a combination of the qualities of news items and published user comments.

Our work in this paper is motivated by the apparent disagreement between the findings from different communities. We aim to understand the factors-- ranging from article content to observed dynamics of user comments -- on predicting the eventual comment volume an article receives. One may notice a problem here. On the one hand, one would like to predict the comment volume before an article's publication. On the other hand, one has access to user comments only after the article has been online for some time. Nonetheless, the number of eventual total comments an article will get remains relevant. Thus, we relax the problem by formulating it as follows:

\vspace{3pt} %vertical space
\noindent\textbf{Problem:} Given a news article $A$ and its first $\alpha$ user comments, predict $N_A$, the eventual number of comments $A$ receives.
\vspace{3pt} %vertical space

We can draw a parallel to the problem of predicting the distance traveled by a ball (news article), say in soccer. The distance depends on the ball itself and the person who kicks it (news outlet and author), but it also depends on factors, like launch angle and exit speed, unrelated to the ball. Those are only known shortly after the ball was kicked and traveled a short distance. Similarly, we expect the first $\alpha$ comments to give us the missing information necessary to predict the eventual number of comments the article receives.

%One may notice that 
One may notice that $N_A$
%, the eventual number of comments article $A$ receives, 
is not well defined: theoretically, it may continue to grow endlessly with time. In practice, however, this does not happen for news articles. We monitored each article for 3 months. The articles accumulated 99.84\% of their overall comment volumes within a week and 99.97\% within a month.
\textit{Consequently, hereafter $N_A$ is the number of comments accumulated in the first week by news article $A$}, which empirically is almost identical to the true total number of comments that $A$ receives in practice.

The magnitude of $\alpha$ and its relation to $N_A$ may trivialize the problem, say, ``look at the first $\alpha$ = 1,000 user comments and predict if the article will receive $N_A$ = 1,050." We study the dynamics among the very first few comments and aim to predict if $N_A$ will reach 1,050. We empirically test $\alpha$ = 5, 10, 15, 20, and 50. The accuracy of prediction increases by about 9\% from $\alpha$ = 5 to $\alpha$ = 10, and by less than 2\% from $\alpha$ = 10 to $\alpha$ = 50. We set $\alpha$ =  10 in all our empirical studies. Thus, we aim to predict the eventual number of comments an article will receive based on the observations among the first 10 comments. It takes 25 minutes on average for the 10$^{\text{th}}$ comment to arrive since the posting of the first comment.

After evaluating and comparing the prediction performance of various models on 19K articles and over 9M comments from 6 news outlets, we show that signals gathered from the early dynamics of user comments largely influence the ability to predict the eventual number of comments on a news article, while the contribution from article features is small. This finding is consistent with the conclusion from the social science community. 
We study the user comment features and identify the feature of ``the arrival rate of \emph{early} comments" (\texttt{rate}), which is defined as the number of comments per minute, as a key missing link in accurately predicting the comment volume.
%feature that precluded the previous work from achieving high prediction accuracy.
%While arrival patterns of user posts are considered in previous works \cite{backstromKLD13,wengMA14}, they are different from  \texttt{rate} as defined in this paper. We elaborate more on this in the Related Work section.
%In addition, those patterns are not as consequential in their respective prediction tasks as \texttt{rate} is in ours. For instance, arrival patterns as defined in \cite{backstromKLD13} contributes less than 4.4\% to the overall performance, compared to 90\% on average for \texttt{rate}.  The family of features ``growth rate" \cite{wengMA14}, which includes a feature similar to \texttt{rate}, has a much weaker predictive power than that of their other features.

%We elevate the study of \texttt{rate} from individual news articles to news outlets and news categories. 
We study \texttt{rate} across six major U.S. and U.K. news outlets. We notice that the performance of the rate-based model varies across news outlets. It is highly accurate for news articles published by Wall Street Journal, but less accurate for Fox News and the Guardian. 
%To understand the nature of the rate model, we present an in-depth study of rate and comment volume in log scale, using fitted linear regressions. 
With regard to the analysis of \texttt{rate} by categories, we note that the characteristic of the rate model differs across categories. For example, ``Politics" is particularly sensitive to the rate of early comments. This is common across all news outlets. ``Health," on the other hand, is less sensitive to the rate of early comments. 

We also consider the relationship between news outlets and categories. We study the characteristic of \texttt{rate} model for each outlet-category pair. We find that the rate model performs the best at Wall Street Journal in most of the categories, 
and the eventual user activity is more affected by the initial engagement in political areas across all outlets.

We believe that our findings are of interest to social scientists because they reveal the relationship between the early user commenting behavior and the total comment volume of a news article, across news outlets and news categories. \textit{This appears to be a trait unique to news readership communities. Features based on early arrival pattern in other social communities, such as Twitter and Facebook, on related prediction tasks have limited predictive power} \cite{backstromKLD13,wengMA14}. Network topology features are more effective in those tasks; most of those features are not applicable to news readership communities as they lack an underlying network.

We make the following contributions in this paper:
\begin{itemize}%[noitemsep, topsep=0pt]
    \item We postulate that one cannot predict the  comment volume of an article unless one considers (early) user commenting activity.
	%\item We identify the importance of user factors, particularly that of \texttt{rate}, in the task of predicting the user comment volume of a news article.
	\item We identify the importance of (early arrival) \texttt{rate} in the task of predicting the comment volume of a news article. 
	\item We perform extensive empirical studies by news outlets and news categories, and show additional novel insights.
	%\item We introduce novel features, some of them are inspired from \emph{news value theory} which capture text level signals (e.g., negativity, positivity, aggression), and others reflect the user commenting behavior (e.g., rate, continuity).
	
\end{itemize}

\section{Related Work}
\label{sec:relatedWork}

%The study of human interaction in online media and social networks has seen increased interest since the advancement of Web 2.0. 
We review several lines of research about user generated content in news domain and social networks.
The problem has been tackled from a variety of angles, such as the relation between comments and the popularity of weblogs \cite{mishneG06}, diversionary role of comments \cite{wangYYLM15}, conversation subjectivity \cite{biyaniBCM12}, and user participation in online forums \cite{delMSWMZ17,manikondaPKHM16,roweA14,schneider2018leveraging}.

%  . \cite{manikondaPKHM16} studies the language metrics in goal-oriented forums and general forums, and observe that users of the former contribute to a more cohesive conversation. \cite{biyaniBCM12} studies the dialogue act and subjectivity clues expressed in comment posts to identify whether a conversation is subjective or non-subjective. \cite{delMSWMZ17} concludes that users restrict their attention to a specific set of sources by exploring the dynamics behind the discussion. \cite{roweA14} defines a common framework for engagement analysis and comparison across five social media platforms. It does not include news outlets.

\textbf{News Domain.} Mining and analyzing the content produced by users in news media are popular research directions. 
Some of the explored problems include 
%collecting and analyzing the comments on news articles \cite{schuthMD07}; 
examining the relationships between news comment topicality, temporality, sentiment, and quality \cite{diakopoulosN11,he2020dynamics,liu2015florin}; analyzing the sentiment of comments and headlines of news article \cite{dosFPRHJ15}; 
news propagation \cite{tanFA16};
%news propagation: from the information source to news article to user comment \cite{tanFA16}, where sentiment is utilized to predict how far the article is from the information source.
personalized recommendation of news stories \cite{shmueliKKL12}; topic clustering of news articles \cite{akerKBPBHG16}; and modeling and predicting comment volume \cite{tsagkiasWD10,balaliAF17,rizosPK16,tatarLALDF11}. 
%\cite{shmueliKKL12} combines the article information and co-commenting patterns of users for personalized recommendation of news stories.
%Topic clustering is another example of a task to which one can give a more accurate solution if one combines news articles and their user comments. The studies in \cite{akerKBPBHG16,Llewellyn16,maSYC12,tanFA16} show that the content of comments and replies combined with the news articles can lead to superior results in the task of topic clustering.
%To better understand the user online interest, some studies focus on modeling or predicting comment number. For instance, one study shows that comments from eight news agencies can be modeled by log-normal and negative binomial distributions \cite{tsagkiasWD10}. 
%On the task of predicting the comment volume of a news article, a previous work \cite{tsagkiasWD09} models it as a two stage classification problem: whether a news article has the potential to receive any comments at all, and if so, whether it will receive comments in high or low volume.
%On the task of modeling the comment volume of a news article, both log-normal and negative binomial distributions have been explored \cite{tsagkiasWD10}. 
The prediction of comment volume is treated as a (binary) classification problem (e.g.,
%``With"\textfractionsolidus{} ``Without" comment, 
``High"\textfractionsolidus{} ``Low" volume) \cite{tsagkiasWD09}
%, multiclass classification (``No comment", ``Moderate", and ``High" ), 
and regression classification problem \cite{balaliAF17} in previous studies. 
\cite{tatarLALDF11} uses a simple linear regression model with early user activity during a short observation period after publication to predict the comment volume of articles. Besides, \cite{aragonGGK17} describes few current models of the growth of comment threads.

\textbf{Social Networking Platforms.} User behaviors in social media and news platform are connected to each other \cite{stanojevic2019biased}. Understanding user behavior in social networking platforms, e.g. Twitter and Facebook, has attracted large interest. Some studies aim to understand user conversations and their evolution over time \cite{wangYH12}, while others study the commenting and comment rating behavior \cite{siersdorferCPAN14}.
%We give a cursory overview to help position our work with this area of work. 
A number of works address problems related to the prediction of reply volume. They employ a variety of features, such as bag of words \cite{yanoS10}, the arrival patterns of early comments \cite{backstromKLD13}, network specific, like ``followship," and historical behavior in retweet \cite{artziPG12}. The prediction problem itself is modeled in a variety of ways. 
Some model it as a binary classification problem \cite{artziPG12,backstromKLD13}.
%Some model it as a binary classification: e.g.,  predict whether a given tweet will get a response \cite{artziPG12} or whether a Facebook reply thread will exceed a threshold of 8 comments, given the state of the first 5 posts \cite{backstromKLD13}. 
Other formulations include regression \cite{tsurR12}, multi-label classification \cite{wengMA14}, cascades size prediction \cite{chengADKJL14,kobayashiL16}, self-exciting point process \cite{mishraRX16}, or Hawkes process modeling \cite{zhaoEHRL15,rizoiuMKCX18}. %\cite{zhaoEHRL15,rizoiuXSCYV17,rizoiuMKCX18,caoSCOC17}.
The popularity prediction and modeling on social media is also a fruitful research \cite{liaoXLHLL19,mishra19,linHLYL19}.
%The popularity prediction and modeling on social media is also a fruitful research: \cite{liaoXLHLL19} explores the deep fusion of temporal process and content features, \cite{mishraRX18,mishra19} propose a recurrent neural network for modeling asynchronous streams, \cite{linHLYL19} presents a Layer-wise Deep Stacking model.

\textbf{Our Work.} 
%Our objective is to understand the factors in accurate prediction of the comment volume a news article receives.
While our work shares some commonalities with these lines of work, it also distinguishes from them in several important ways. 
The main difference with the work in the news domain is that we focus on the analysis of (early) user commenting activity and its importance on the prediction of comment volume of a news article.
The study of using user comments in early stage to predict the final comment volume has been analyzed in \cite{tatarLALDF11}, which only evaluates a simple linear prediction model with limited factors from articles and comments. While our work in this paper explores more features related to article and early user commenting activity (both new and old). 
Moreover, we consider multiple machine learning techniques, both linear and nonlinear. According to our analysis, we show that (1) the prediction problem is difficult and, thus, nonlinear models are better suited to solve the prediction problem; and (2) the proposed new feature \texttt{rate} remains its dominant power across machine learning techniques.
%we treat the prediction task as a regression task instead of a binary classification task and we analyze the role of user commenting factors, e.g., commenting rate, in the prediction of the user comment volume. 
The key distinction with the work in social networks is that the social communities at news outlets are \textit{not} networked. 
%consist of people with no explicit relation among them.
%, e.g., Fox News hosts a social community where the majority of people share conservative views and there is no explicit link between them. 
The works in social networks make heavy use of the network topology and the community around a user, e.g., followers and friends. These are not applicable in our setting. 
%\textcolor{blue}{While arrival patterns of user posts are considered in previous works, they are different from \texttt{rate} as defined in this paper, to our knowledge.} In addition, those patterns are not as consequential in their respective prediction tasks as \texttt{rate} is in ours. 
While arrival patterns of user posts are considered in previous works, they are not as consequential in their respective prediction tasks as \texttt{rate} is in ours.
For instance, arrival patterns as defined in \cite{backstromKLD13} contribute less than 4.4\% to the overall performance, compared to 90\% on average for \texttt{rate}.  The family of features ``growth rate" \cite{wengMA14}, which includes a feature similar to \texttt{rate}, has a much weaker predictive power than that of their other features. Since their \texttt{rate}s are inconsequential, these works do not pursue any in depth studies of their \texttt{rates}. We present a study of \texttt{rate} along several dimensions, such as news outlet and news category.

\begin{figure}[!t]
	\centering
	%\vspace{-2mm}
	\includegraphics[width=0.95\linewidth, height=0.48\linewidth]{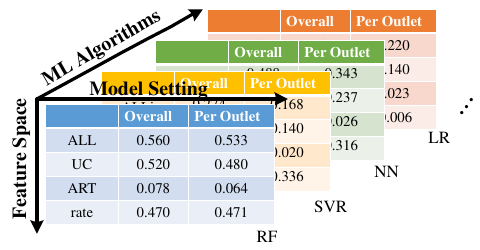}	
	\vspace{-10pt}
	\caption{Methodology illustration.}
	\label{fig:methodologyCartoon}
	\vspace{-5pt}
\end{figure}

\section{Methodology}
\label{sec:methodology}
Our goal is to understand the feature subset most important for predicting the comment volume of a news article. The prediction of the eventual comment volume is a regression problem. Figure \ref{fig:methodologyCartoon} summarizes our methodology. We conduct our study along three dimensions: (i) Feature Space, (ii) Model Setting, and (iii) Machine Learning (ML) Algorithms. In (i), we analyze the entire feature space (denoted as ALL), the user comment only (UC) and news article only (ART) features, as well as \texttt{rate} (which is a single feature in UC) alone. 
%ACOMi uses all features listed in our paper. COMi contains comment features only. ART resembles the model described in \cite{tsagkiasWD09} by utilizing article related features, which is treated as the baseline. Rate means that only the feature \texttt{rate} is included in the feature space. 
In (ii), we consider two settings: \textit{global} and \textit{local}. The global dataset has the news articles from all the news outlets. The local dataset has the news articles grouped by news outlet. % Then, we train each of ACOMi, COMi, ART, and Rate on the  overall dataset. For the local one, we train and test the models on news articles within each outlet. The test data is disjoint from the train data in all scenarios.
In (iii), we use 4 representative ML algorithms for regression: Random Forest (RF), Support Vector Regression (SVR), Neural Network (NN), and Linear Regression (LR). In the figure, a slice represents an instance from the cross product of (i), (ii), and (iii).
%we repeat the experiments based on different feature space and model setting. The results from the same ML method are displayed in the same slice. With more ML methods added, there would be more such kind of slices.

\begin{figure*}[t!]
	\centering
	%\begin{subfigure}[h]{1.0\textwidth}
	%	\centering
	%	\includegraphics[width=0.33\linewidth, height=0.16\linewidth]{volumeWSP}
	%	\includegraphics[width=0.33\linewidth, height=0.16\linewidth]{logVol_wsp}
	%	\includegraphics[width=0.3\linewidth, height=0.16\linewidth]{wspYQQ}
	%	\caption{Washington Post}
	%	\label{fig:volumeWSP}
	%\end{subfigure}
	\begin{subfigure}[h]{1.0\textwidth}
		\centering
		\includegraphics[width=0.33\linewidth, height=0.16\linewidth]{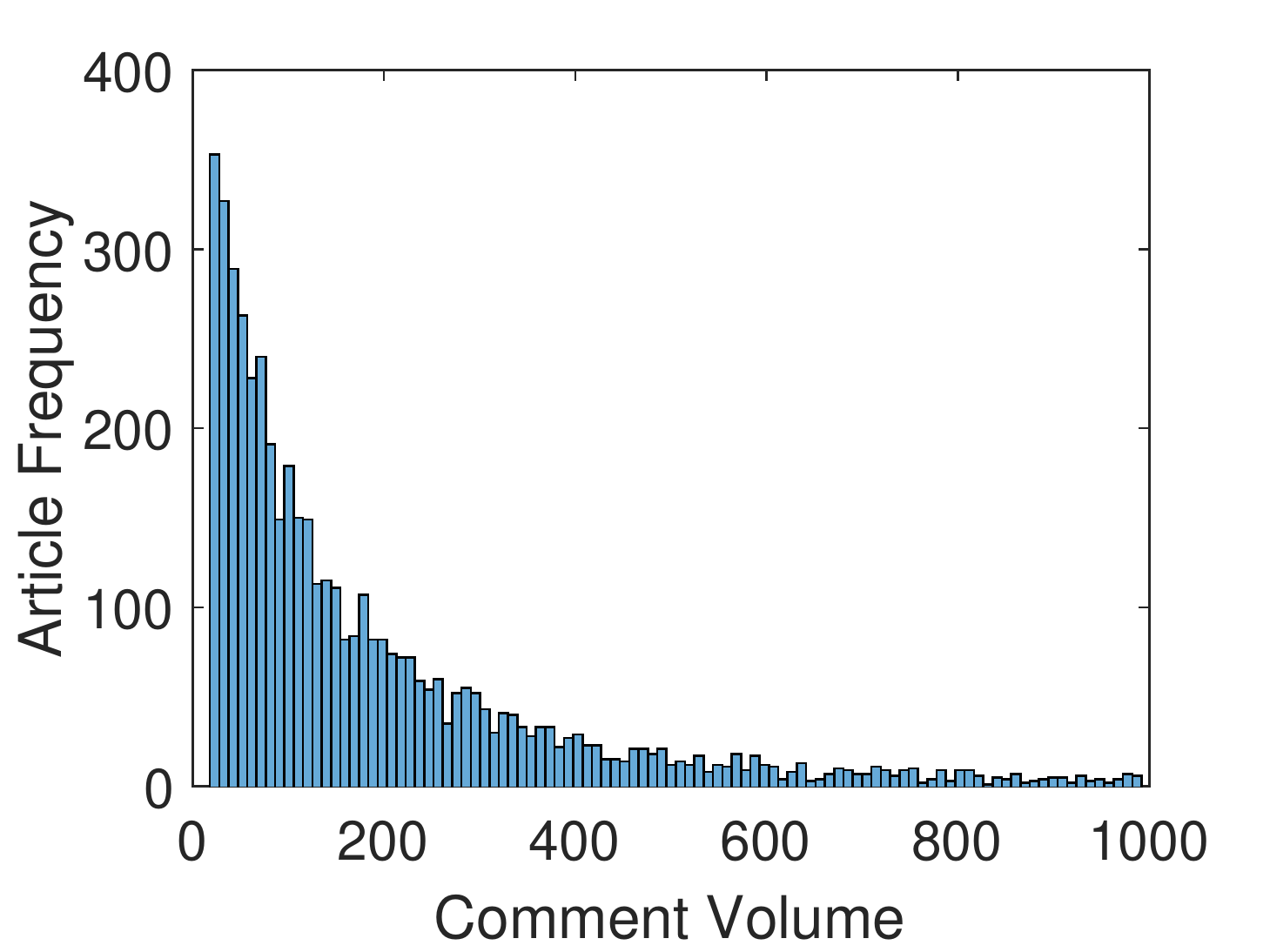}
		\includegraphics[width=0.33\linewidth, height=0.16\linewidth]{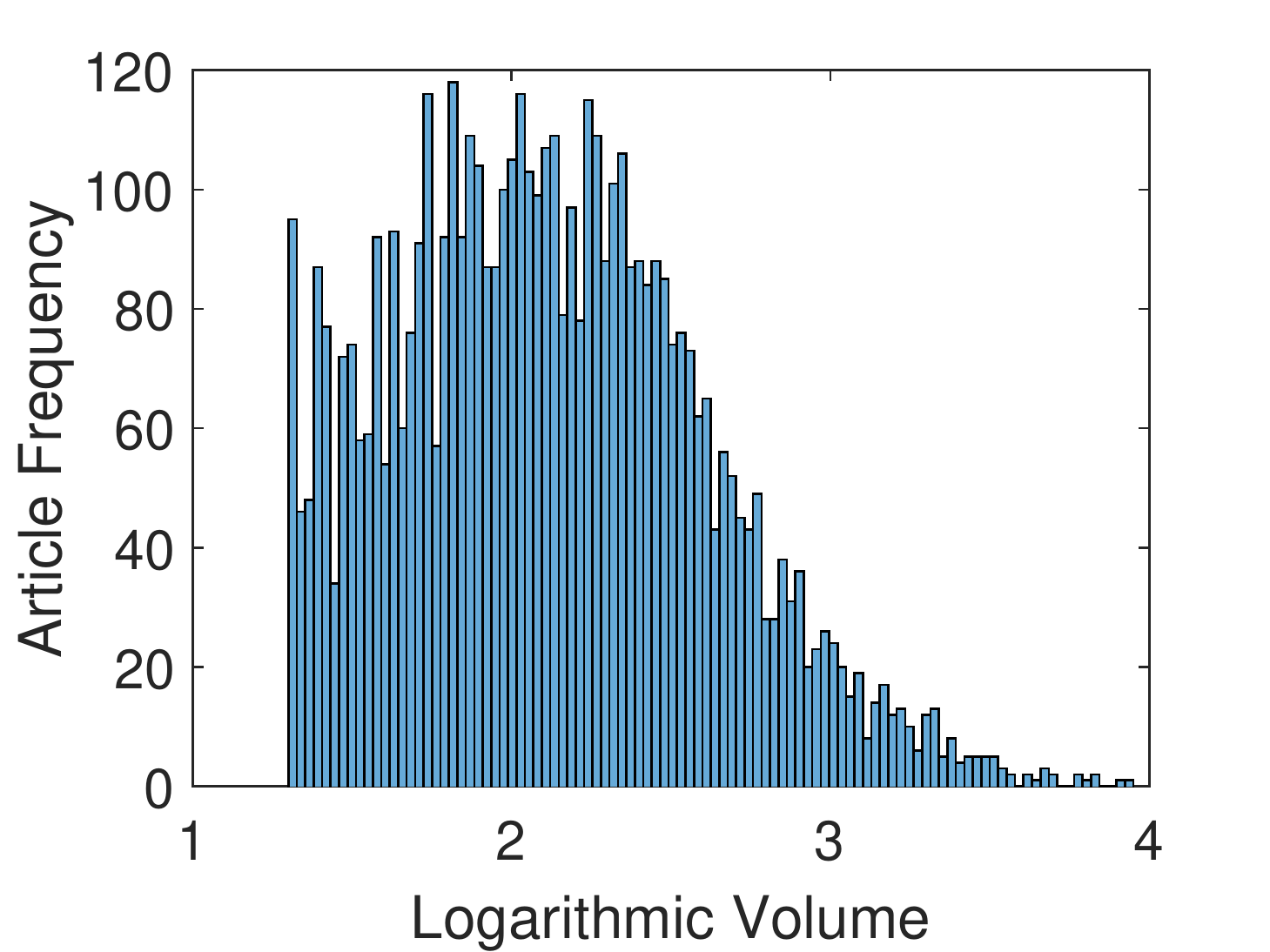}
		\includegraphics[width=0.3\linewidth, height=0.16\linewidth]{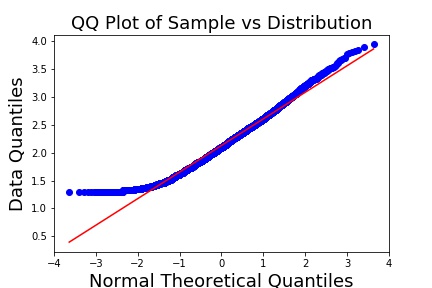}
		\vspace{-4pt}
		\caption{Daily Mail}
		\label{fig:volumeDM}
	\end{subfigure}
	%\begin{subfigure}[h]{1.0\textwidth}
	%	\centering
	%	\includegraphics[width=0.33\linewidth, height=0.16\linewidth]{volumeWSJ}
	%	\includegraphics[width=0.33\linewidth, height=0.16\linewidth]{logVol_wsj}
	%	\includegraphics[width=0.3\linewidth, height=0.16\linewidth]{wsjYQQ}
	%	\caption{Wall Street Journal}		
	%	\label{fig:volumeWSJ}
	%\end{subfigure}	
	%\begin{subfigure}[h]{1.0\textwidth}
	%	\centering
	%	\includegraphics[width=0.33\linewidth, height=0.16\linewidth]{volumeFox}
	%	\includegraphics[width=0.33\linewidth, height=0.16\linewidth]{logVol_fox}	
	%	\includegraphics[width=0.3\linewidth, height=0.16\linewidth]{foxYQQ}
	%	\vspace{-4pt}
	%	\caption{Fox News}					
	%	\label{fig:volumeFox}	
	%\end{subfigure}
	\begin{subfigure}[h]{1.0\textwidth}
		\centering
		\includegraphics[width=0.33\linewidth, height=0.16\linewidth]{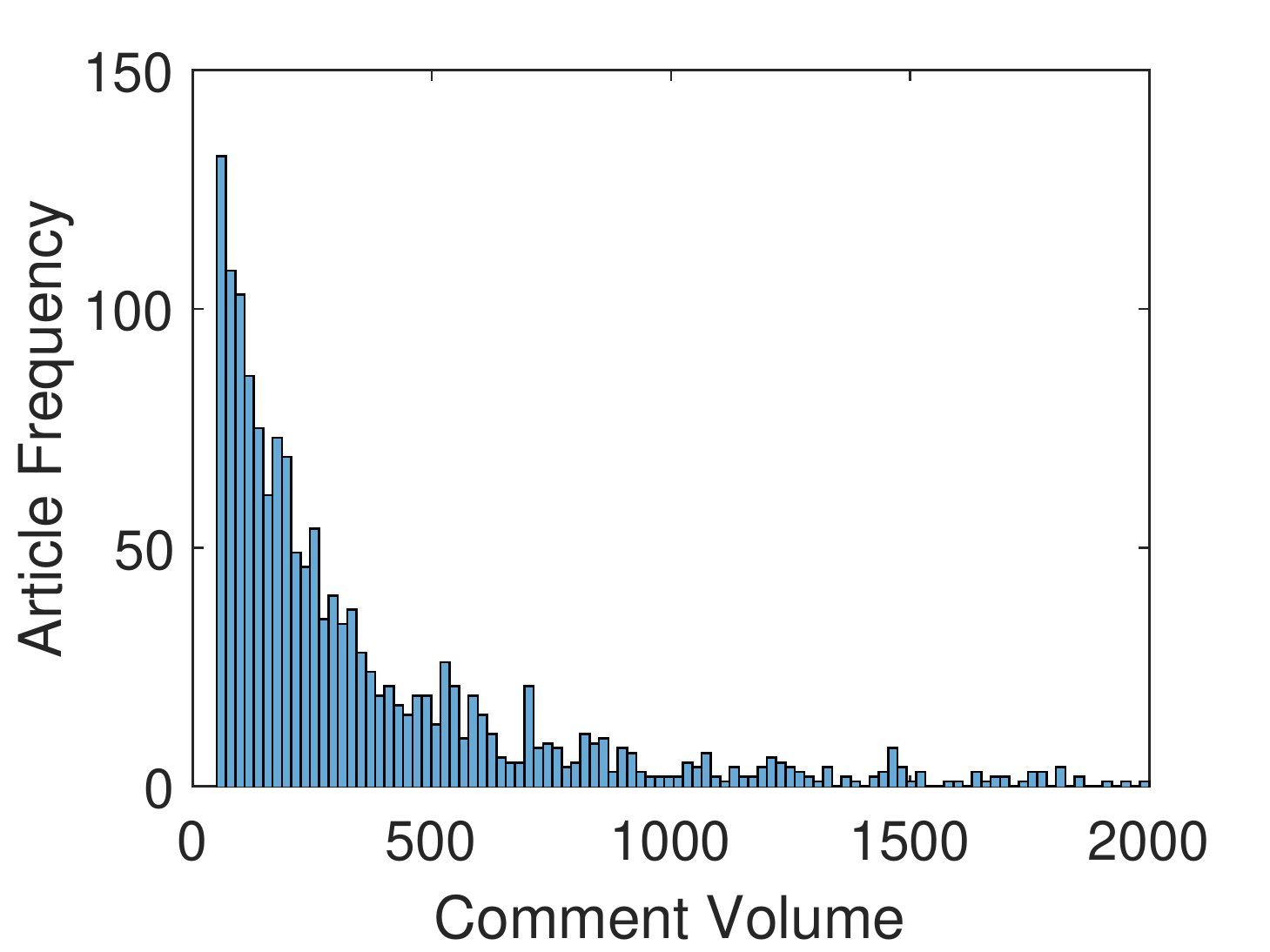}
		\includegraphics[width=0.33\linewidth, height=0.16\linewidth]{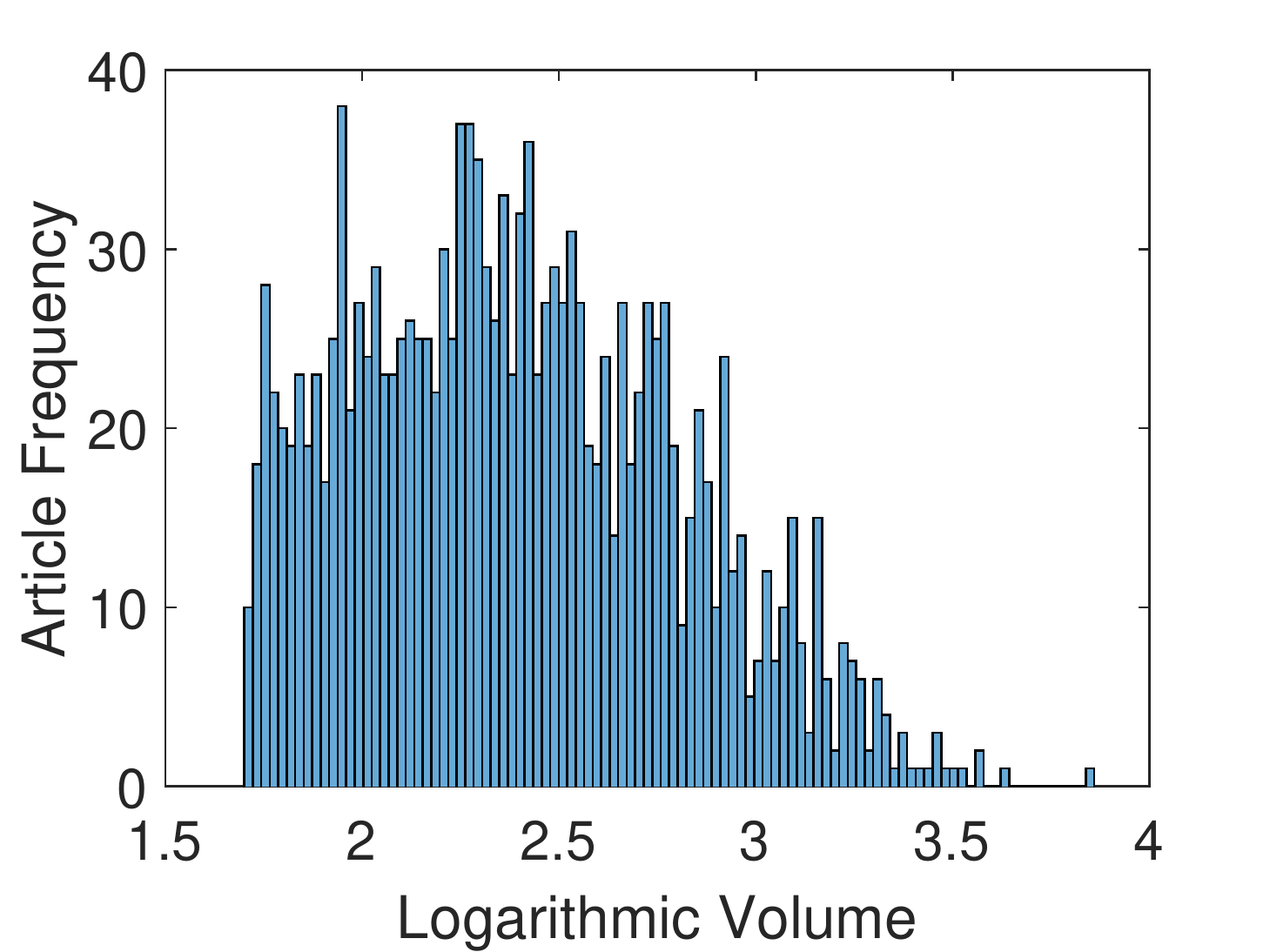}
		\includegraphics[width=0.3\linewidth, height=0.16\linewidth]{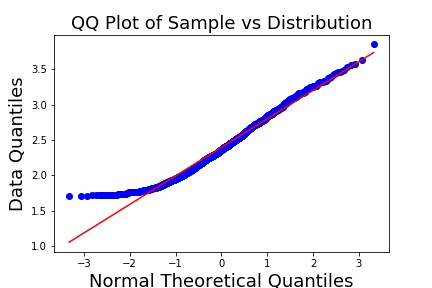}
		\vspace{-4pt}
		\caption{the Guardian}		
		\label{fig:volumeGuardian}
	\end{subfigure}
	%\begin{subfigure}[h]{1.0\textwidth}
	%	\centering
	%	\includegraphics[width=0.33\linewidth, height=0.16\linewidth]{volumeNYTimes}
	%	\includegraphics[width=0.33\linewidth, height=0.16\linewidth]{logVol_nyt}	
	%	\includegraphics[width=0.3\linewidth, height=0.16\linewidth]{nytYQQ}
	%	\caption{New York Times}
	%	\label{fig:volumeNYTimes}
	%\end{subfigure}
	\vspace{-10pt}
	\caption{The distribution of comment volume and logarithmic volume per news outlet. In the first column of graphs, the articles with more than 1,000/2,000 comments are discarded to make the graphs visible. For each outlet, article frequency on the y-axis of the first two graphs is the number of articles, the third graph provide the Q-Q plot of the logarithmic volume.}
	\label{fig:volumeDistribution}
	\vspace{-5pt}
\end{figure*}

\section{Data}
\label{sec:dataPreparation}

%We develop a crawler for news article and comment. It crawls
%We crawled all the articles that appear in Top Stories in Google News from Oct. 2015 to Feb. 2017. 
We collected news articles with comments from Oct. 2015 to Feb. 2017 from the following six news outlets: Washington Post, Daily Mail, Wall Street Journal, Fox News, the Guardian, and New York Times. 
We crawled the topics of the collected articles from Google News and monitored their duration there as well.
%We noticed that not all crawled articles received comments (see the columns Articles and Aw.C in Table \ref{tab:summary}). The main reason for an article not receiving comments was that the news outlet did not make it available for commenting (for reasons unknown to us). 
The dataset has over 19K articles with comments and 9M comments (including replies). 
%Hence, our dataset has about 19K articles with comments, which belong to 768 distinct topics. For the collection of comments, a reply to a previous comment is counted as a comment as well.
We monitored each article for 3 months. We observed that on average each article accumulates 99.84\% of its overall comment volume within a week. 
%and 99.97\% within a month. 
%We restate our assumption 
Recall that in this paper $N_A$ is the number of comments accumulated in the first week by news article $A$. This is the number we try to predict.
Figure \ref{fig:volumeDistribution} illustrates the distribution of news articles at these outlets in our dataset (we give 2 outlets due to space constraints). We observe a heavy-tailed distribution in each outlet when we plot the distribution by number of comments (the first graph per outlet). If we plot the comment volume in the log scale, the distributions are (or close to be) bell-shaped. It seems that the volume distributions are nearly log-normal.
To test this hypothesis, we provide the Q-Q plot of the logarithmic volume as the third column of graphs in Figure \ref{fig:volumeDistribution} for each outlet. We can see that most of the points stay on or very close to the straight line, except for some head and tail data points. It shows that the distribution of user comment volume over news articles is well approximated by the log-normal distribution.

Table \ref{tab:summary} describes the number of articles in each news outlet. 
We give the mean and standard deviation (STD) of comment volume and logarithmic volume, displayed in the last two columns of Table \ref{tab:summary}. 
%It shows that the logarithmic volume does not fluctuate as drastically as comment volume by comparing the STDs to the mean in each outlet.
Considering the log-normal distribution of comment volume,
%and the relatively small fluctuation of logarithmic volume, 
we will work on the prediction of comment volume in log scale.
%both in the prediction task and the in-depth study of feature \texttt{rate}.

%A number of features related to news factors \cite{weber14} (e.g., continuity) require (meta)data beyond that of a news article and its comments. For example, we gather news category for each news article from both the news outlet and Google News. The latter requires that we continuously crawl the News Topics in Google News.

\begin{table}[t]
	\small
	\centering
	\caption{Data summary. Aw.C = Articles with Comments.}
	\label{tab:summary}
	\vspace{-3mm}
	\begin{tabular}{|@{}c@{}|c|@{}c@{}|@{}c@{}|}
		\hline
		\textbf{Outlets}  & \textbf{Aw.C} & \tabincell{c}{\textbf{Mean Vol.} \\\textbf{(STD)}} & \tabincell{c}{\textbf{Mean Log Vol.}\\\textbf{(STD)}}\\
		\hline		
		Washington Post & 6,470 & \tabincell{c}{364.8 (942.2)} & \tabincell{c}{1.88 (0.76)}\\
		\hline
		Daily Mail  & 6,046 & \tabincell{c}{264.1  (560.4)} & \tabincell{c}{1.99 (0.62)}\\
		\hline
		Wall Street Journal & 2,516 & \tabincell{c}{189.4 (346.5)} & \tabincell{c}{1.74 (0.7)}\\
		\hline
		Fox News & 1,739 & \tabincell{c}{1,896.5  (3790.2)} & \tabincell{c}{2.47 (0.94)}\\
		\hline
		the Guardian  & 1,697 & \tabincell{c}{504.4 (716.8)} & \tabincell{c}{2.46 (0.45)}\\
		\hline
		New York Times  & 965 & \tabincell{c}{481.4 (530.9)} & \tabincell{c}{2.38 (0.6)}\\
		\hline
		Overall  & 19,433 & \tabincell{c}{465.8 (1400.6)} & \tabincell{c}{2.02 (0.74)}\\
		\hline
	\end{tabular}
	\vspace{-5pt}
\end{table}

\section{Predicting Comment Volume}
\label{sec:predictVolume}

In this section, we show that factors drawn from (early) user commenting activity 
%, and in particular \texttt{rate}, 
are the keys to accurately predict the comment volume 
%in the log scale 
a news article receives. We describe the feature set and the experimental setting, and report on the prediction performance in this section. 
We follow the methodology described above. 
Finally, we demonstrate that \texttt{rate} is the dominant feature.

\subsection{Features}

\begin{table}[!t]
	\small
	\centering
	\caption{Features utilized in prediction experiments. We organize them into 5 groups. The underlined ones belong to two groups, one of which is news factors. The features labeled with * are new features in the prediction task.}
	\label{tab:features}
	\vspace{-2mm}
	\begin{tabular}{|l p{5.0cm}|}
		\hline
		\textbf{Feature} & \textbf{Description}\\
		\hline
		\textbf{Topic features} &  \\
		\hline
		topic* & Topic of article. \\
		\hline
		\textbf{Article features} & \\
		\hline
		\underline{month}  &  Published month of article (1-12). \\
		\underline{day}    &  Published day of the month (1-31). \\
		\underline{hour}   &  Published hour of the day (0-23). \\
		wom    &  Week of the month (1-5). \\
		dow    &  Day of the week (1-7). \\
		author &  Author of article. \\
		\underline{art\_length}  &  Article content length. \\
		\underline{art\_question}     &  Whether there is a '?' in article title. \\
		art\_exclaim      &  Whether there is a '!' in article title. \\
		art\_num\_ne\_loc &  Number of location-type named entities in article content.
		\\
		art\_num\_ne\_per &  Number of person-type named entities in article content.
		\\
		art\_num\_ne\_org &  Number of org.-type named entities in article content.
		\\
		art\_num\_ne\_misc &  Number of miscellaneous-type named entity in article content. \\
		\underline{art\_senti\_score*} & Sentiment score of article content. \\
		\hline	
		\tabincell{c}{\textbf{Comment features}}  & \\
		\hline
		rate*    &  Arriving rate of the first $\alpha$ comments.\\	
		fc\_mid* & Time of first comment - 12am (in min.) \\
		uniq\_com  & Number of unique commenters. \\
		num\_reply  &  Number of replies. \\
		num\_thread &  Number of threads. \\
		\underline{num\_question} & Number of '?'. \\
		num\_exclaim  & Number of '!'. \\
		\underline{num\_words}  & Number of words. \\
		complexity & Complexity of the first $\alpha$ comments. \\
		\underline{has\_url}   & Whether there is a link. \\
		num\_ne\_com & Number of named entities. \\
		depth*       &  Depth of the comment tree. \\
		width*       &  Width of the comment tree. \\
		\underline{avg\_senti\_score*} & Average of sentiment scores of the first $\alpha$ comments. \\
		num\_likes & Aggregated number of likes. \\
		num\_dislikes & Aggregated number of dislikes. \\
		\hline
		\textbf{News Factors} & \\
		\hline		
		continuity*  &  Time difference between article's publication and its topic's appearance. \\
		%proximity  & Whether the country where the event took place is same with the news outlet.\\
		%negativity*  & Fraction of negative words.\\
		%positivity*  & Fraction of positive words.\\
		aggression*  & Fraction of aggressive words.\\	
		position    & 	NA. \\
		\hline
		\textbf{MISC features} & \\
		\hline
		pub\_resp   &  Time difference (in minutes) of first comment to article's publication. \\
		inter\_art*    &  Defined as $ \frac{|NE_{art}\bigcap NE_{com_i}|}{|NE_{art}|}$ \\
		inter\_com*    &  Defined as $\frac{|NE_{art}\bigcap NE_{com_i}|}{|NE_{com_i}|}$ \\ 		
		\hline
	\end{tabular}
	\vspace{-5pt}
\end{table}

Table \ref{tab:features} summaries the set of features. There are five groups of features: topic, article, comment, news factors, and misc features. We introduce 11 new features.
%spread across the five groups.

\textit{\textbf{Topic features}.} We observe that some topics, such as Ebola Outbreak or Paris (terrorist attack), trigger more discussion than others. Therefore, we include these finer grain topics as one of the predictive features. The fine grain topics are rarely provided by the news outlets. We extract them from Google News along with their parent categories, e.g., Health and World.
We collect 768 distinct topics in total.

%Each news article is on finer grain topic, such as Paris (terrorist attack) or Ebola Outbreak.

\textit{\textbf{Article features}.}  All features in this group are related to news articles. They can be categorized into metadata and text features. The metadata features include \textit{month}, \textit{day}, \textit{hour}, \textit{wom} (week of the month), and \textit{dow} (day of the week) of the publication. Previous work argues that the time of publication may affect the comment volume an article receives \cite{tsagkiasWD10}. The rest of the features in this group are extracted from article title and content. The features \textit{art\_question} and \textit{art\_exclaim}, suggested in \cite{backstromKLD13}, show whether there are '?' and '!' in article title.
The features \textit{art\_num\_ne\_loc}, \textit{art\_num\_ne\_per}, \textit{art\_num\_ne\_org}, and \textit{art\_num\_ne\_misc}, proposed in \cite{tsagkiasWD09}, provide the number of locations, persons, organizations, and miscellaneous named entities mentioned in the article content. Previous work \cite{zhang2018regular,zhang2019invest} proposed some possible methods to extract these named entities. In this paper, we utilize Stanford NER
%Named Entity Recognizer 
to extract named entities. 
%These features consider the semantic influence of an article.

The feature \textit{art\_senti\_score} gives the sentiment score of article content. Sentiment lexicons, studied in \cite{dragut2010construction,schneider2015towards,schneider2018debugsl}, are possible ways to calculate this score. In this paper, we make use of an effective document representation and sentiment analysis model proposed in \cite{yangYDHSH16}, which is a word-sentence-document level bi-directional GRU neural network with two levels of attention. We initialize the 100 dimensional word embeddings with pre-trained Glove word vectors. The model is trained on IMDB dataset (25K reviews with positive or negative rates) for 10 epochs. We use the output from the prediction layer of the deep model as sentiment score, which is in the range of [0, 1]. Articles with score close to 0 are predicted with overall negative sentiment, while articles with score close to 1 are predicted with overall positive sentiment. 

\textit{\textbf{Comment features}.} The comment features are extracted from the first $\alpha$ comments of an article. 
%They can be categorized into temporal dynamics, text, and reply tree structure. 
The feature \textit{rate} measures the number of comments per unit of time, which is computed as 
$$rate=\displaystyle\frac{i}{t_i-t_1}$$
Here, $t_i - t_1$ is the elapsed time (in minutes) between the first and \textit{i}-th comment (as in \cite{wengMA14}); it is 25 minutes on average for the 10$^{\text{th}}$ comment. The feature \textit{fc\_mid} is the absolute difference between the time of the first comment and midnight.
%These four features reflect the user commenting behavior at the front of the commenting stream.
%As implied in \cite{tsagkiasWD10}, early commenting activity is predictive of the final volume of comments.
The feature \textit{uniq\_com} gives the number of unique commenters in the first $\alpha$ comments, which is one of the indicators for the arrival pattern of the first $\alpha$ comments \cite{backstromKLD13}. The features \textit{num\_reply} and \textit{num\_thread} give the number of replies and discussion threads, respectively. %Previous studies show that there are few types of dialogue acts expressed in comment threads because of their conversational nature \cite{biyaniBCM12,bhatiaBM12,jotyCL11,jeongLL09}. Question mark is the typical one.

The features \textit{num\_question}, \textit{num\_exclaim}, \textit{num\_words}, \textit{complexity}, \textit{has\_url}, \textit{num\_ne\_com}, and \textit{avg\_senti\_score} study the text of comments. The meaning of these features are provided in Table \ref{tab:features}.
\textit{Complexity} measures the cumulative entropy of terms within the first $\alpha$ comments \cite{roweA14}. It is given by:
\[
complexity(c) = \textstyle \frac{1}{|T(c)|} \sum\limits_{t \in T(c)} tf(t,c)(\log|T(c)| - \log tf(t,c)) 
\]
\setlength{\belowdisplayskip}{1pt}
Here, $T(c)$ is the set of unique terms in comment \textit{c} and $tf(t, c)$ is the frequency of each term $t\in T(c)$.
%It returns a high value if the comment contains many non-repeated terms. This measure can distinguish two comments which have the same word count but one of them is generated by repeating some words.
%such as ``TRUMP TRUMP TRUMP" and ``Mashed potatoes! Trumpy!"\footnote{Two comments from the article \url{http://www.dailymail.co.uk/news/article-3385411/The-candidate-beat-Trump-crushing-Cruz-nationally-four-weeks-2016-contest.html}}.

The feature \textit{avg\_senti\_score} is the average of the sentiment scores of the first $\alpha$ comments. The sentiment score of a comment is given by the deep model in \cite{yangYDHSH16}.
%, same as how we achieve \textit{art\_senti\_score}.}

\begin{comment}
\begin{figure}[t]
	\centering
	\includegraphics[width=0.7\linewidth]{commentTree.png}
	%\vspace{-3mm}
	\caption{Comment conversation tree $T_A$.}
	\label{fig:commentTree}	
	%\vspace{-10pt}
\end{figure}
\end{comment}

The features \textit{depth} and \textit{width} are extracted from the comment reply tree $T_A$ of an article $A$. $T_A$ is constructed as follows.
%in Figure \ref{fig:commentTree}.
An article $A$ is the root of $T_A$. Comments that are not replies (responses) of any previous comments are the children of $A$ (the article). The replies of a comment are its children nodes. The \textit{depth} of the reply tree $T_A$ is the number of levels of $T_A$.
%, denoted $L1, L2$ in the figure.
%The width is calculated from the maximum of siblings summation in each level, which is
If $L$ denotes the levels of the reply tree $T_A$, the \textit{width} is given by
$$ WIDTH = \max \limits_{j\in \textbf{L}} \sum_{i=1}^{m_j} s_{ji}, $$
%\begin{equation} \label{eq:width}
%\begin{aligned}
%WIDTH = \max \limits_{j\in \textbf{L}} \sum_{i=1}^{m_j} s_{ji}
%\end{aligned}
%\end{equation}
where $m_j$ is the number of sibling groups in level $j$, and $s_{ji}$ is the count of nodes in the $i$-th sibling group in level $j$. A feature named \textit{depth} appears in \cite{chengADKJL14}, but its definition and meaning are different from ours.

The features \textit{num\_likes} and \textit{num\_dislikes} count the aggregated number of likes and dislikes received by the first $\alpha$ comments. The consideration of number of likes is proposed in \cite{backstromKLD13}.

\textit{\textbf{News factors}.}  We implement a number of novel features based upon \emph{news value theory}, which states that journalists and media users select news items depending on news factors such as continuity, negativity, and aggression \cite{weber14,ZiegeleQ13}. These dimensions were confirmed after extensive face-to-face interviews with users who commented on news stories online \cite{ZiegeleBQ14}. We create novel features to quantify many of the news factors. Some of them are encountered in other studies, e.g.,  climate change \cite{olteanuCDA15}, but with different definitions.
We quantify the factor \textit{continuity} (if a news article continues issues that are already on the media agenda) \cite{weber14} as the time difference (in minutes) between article's publication time and its topic's appearance in Google News. The intuition is that a user's interest to comment on an article diminishes the farther its publication time is from the time when the news event first broke in.
%For \textit{proximity}, the measure is a little different among outlets. For Washington Post, Wall Street Journal, Fox News, and New York Times, we regard articles as proximate if they are about US news. For Daily Mail and The Guardian, however, proximate articles are those whose topics are related to the United Kingdom.
We additionally consider the factors \textit{negativity} and \textit{aggression} both in the article text and user comments. 
To quantify \textit{negativity}, we calculate the sentiment score of a piece of text (article or comment) by applying an effective document representation and sentiment analysis model proposed in \cite{yangYDHSH16}. A piece of text with sentiment score closer to 0 shows stronger negativity, while a text with score closer to 1 indicates stronger positivity. For the sentiment of article and comments, we propose features \textit{art\_senti\_score} and \textit{avg\_senti\_score}, which are already present in the group of article and comment features, respectively.
We use the lexicon LIWC \cite{TausczikP10} to quantify \textit{aggression}. Given a piece of text, \textit{aggression} is defined as the count of entries in the category ``Hostile" together with the ones under ``anger" that appear in the text.
Additional news factors, such as \textit{time of publication}, \textit{uncertainty}, \textit{length}, and \textit{facticity}, are considered in the previous feature groups. They are underlined in the table. Following their definitions \cite{ZiegeleBQ14}, \textit{uncertainty} is measured by the count of question marks in a piece of text and \textit{facticity} is a binary feature, which is 1 if the piece of text contains an URL, and 0 otherwise. The factor \textit{position} of comment in the discussion thread is not applicable in our case, since we only analyze the first $\alpha$ comments.

\textit{\textbf{MISC features}.}
%Many other features have been proposed to predict comment volumes that do not belong to the previous groups and we adopt some of them.
The feature \textit{pub\_resp} describes how fast users respond to an article, which is similar to some of the features in \cite{backstromKLD13,chengADKJL14}. Some works argue that the longer users delay their response to an article, the less overall user activity the article receives \cite{backstromKLD13}. The features \textit{inter\_art} and \textit{inter\_com} quantify the ratio of overlap between the sets of named entities in an article and its first $\alpha$ comments. Let $NE_{art}$ and $NE_{com_i}$ be the sets of named entities that appear in an article and its first $\alpha$ comments, respectively. We define \textit{inter\_art} and \textit{inter\_com} as
$$inter\_art = \frac{|NE_{art}\bigcap NE_{com_i}|}{|NE_{art}|}$$
$$inter\_com = \frac{|NE_{art}\bigcap NE_{com_i}|}{|NE_{com_i}|}. $$

\begin{comment}

$$inter\_art = \displaystyle\frac{|NE_{art}\bigcap NE_{com_i}|}{|NE_{art}|}$$ and 
$$inter\_com = \displaystyle\frac{|NE_{art}\bigcap NE_{com_i}|}{|NE_{com_i}|}.$$
%We borrow additional features from social networking literature.
\end{comment}

\subsection{Experimental Setup}
The experimental study employs the cross-validation methodology. We split the dataset into five folds randomly. The training set consists of articles in four folds. The articles in the remaining fold are used for testing. For a given set of features, we build a model based on the training set, and apply it on a disjoint testing set. The process is repeated five times, each time selecting a different fold for testing. We report the average performance.

\subsubsection{\textbf{Evaluation Metrics}}

\begin{comment}
%Larger value of $R^2$ indicates better result.
It is defined as
%\setlength{\abovedisplayskip}{1pt}
\[ 
R^2 = 1 - \dfrac{\sum_{i=1}^{n} (y_i - \hat{y_i})^2}{\sum_{i=1}^{n} (y_i - \bar{y})^2} = 1 - \dfrac{MSE}{Variance} 
\]

%\setlength{\belowdisplayskip}{1pt}
where, $y_i$ is the observed target value and $\hat{y_i}$ is the predicted target value for object $i$, $\bar{y}$ is the mean of the target values.
%As we can see from Table \ref{tab:summary}, the comment volume differs greatly across news media. It is meaningless to calculate the $R^2$ of the predicted volume directly, especially in the global model ($R^2$ hovers 0.092, showing poor performance in each model). Therefore, we predict the logarithmic comment volume instead of the direct one to mitigate the influence of the volume gap across sources.

\end{comment}

We treat the task of predicting the comment volume of a news article as a regression problem.
%We evaluate each model in the experiments based on 
%$R^2 = 1 - \dfrac{\sum_{i=1}^{n} (y_i - \hat{y_i})^2}{\sum_{i=1}^{n} (y_i - \bar{y})^2}$ and $MAE = \frac{1}{n} \sum_{i=1}^{n} |y_i - \hat{y_i}|$.
We evaluate each model in the experiments based on $R^2$ and the mean absolute error (MAE), which are defined as
\[ 
R^2 = 1 - \dfrac{MSE}{Variance} = 1 - \dfrac{\sum_{i=1}^{n} (y_i - \hat{y_i})^2}{\sum_{i=1}^{n} (y_i - \bar{y})^2} 
\]
\[
MAE = \frac{1}{n} \sum_{i=1}^{n} |y_i - \hat{y_i}|
\]

We calculate the MAE instead of MAPE (mean absolute percentage error) since MAE is more robust to outliers (the long tail in the first graph per outlet in Figure \ref{fig:volumeDistribution}).
Target variable $y_i$ in the calculation of $R^2$ and MAE is the logarithm of the number of comments because the distribution of comment volumes resembles lognormal distribution, as shown in Figure \ref{fig:volumeDistribution}. 
%Another reason is that the standard deviation of the logarithmic volume in each news outlet is much smaller compared to that of comment volume (see STDs in Table \ref{tab:summary}). 
%Therefore, our evaluation metrics, $R^2$ and MAE, are based on the prediction of logarithmic comment volume in this paper.

\subsubsection{\textbf{Hyperparameter Tuning and Setting}}
We consider the first $\alpha$ = 10 user comments for each article when we compute the comment features.
%We remove the articles which have less than 10 comments.
We reached $\alpha$ = 10 after we studied the variation in prediction accuracy for $\alpha$ = 5, 10, 15, 20, and 50, respectively.
The accuracy of prediction increases by about 9\% from $\alpha$ = 5 to $\alpha$ = 10, and by less than 2\% from $\alpha$ = 10 to $\alpha$ = 50. Therefore, we set $\alpha$ =  10 in all our empirical studies.
%We find that the $R^2$ improves only marginally for $\alpha > 10$.
%We employ multiple ML algorithms for the prediction task, as listed in Figure \ref{fig:methodologyCartoon}, to unequivocally show that the dominance of \texttt{rate} is not algorithm dependent. 
We explore multiple machine learning (ML) algorithms for performance comparisons on the proposed comment volume prediction task as listed in Figure \ref{fig:methodologyCartoon}. The results indicate that the feature \texttt{rate} does consistently well across the board, thereby indicating it to be a strong algorithm independent feature that inherently captures the prediction task.

We give a brief overview of the hyperparameter setting for the three nonlinear ML algorithms in our methodology: Random Forest (RF), Support Vector Regression (SVR), and Neural Network (NN). We tune the number of trees (\texttt{ntrees}) for RF. We use SVR with kernel 'rbf' and tune the hyperparameters \texttt{C} and \texttt{$\epsilon$}. We choose Multi-layer Perceptron to implement the NN and tune the hidden layer sizes (\texttt{hsize}) and the initial learning rate (\texttt{lr}); the activation function for the hidden layer is set to be 'relu'. The choices for these hyperparameters are drawn from: $ntrees\in[50, 100, 200, 300]$, $C\in[0.1, 0.5, 1, 5, 10]$, $\epsilon\in[0.01, 0.05, 0.1, 0.5]$, $hsize\in[10, 20, 30, 50, 100, 200]$, and $lr\in[0.001, 0.005, 0.01, 0.05, 0.1]$.
%The number of trees in the Random Forest is 100. We  experimented with other values up to 300, but we did not observe noticeable accuracy gain. The remaining parameters are set to their default values in the literature. For example, the minimum number of observations per tree leaf is 5 and one third of the features are selected at random in each decision split.

\subsection{Experimental Results}
\begin{table}[t]
	\small
	\centering
	%\caption{Comparison of $R^2$ (the first value)/$MAE$ (the second) results based on the overall dataset.		ART is the baseline for each ML algorithm. 		The highlighted row (ART) displays the outcome according to the baselines.}
	\caption{Comparison of $R^2$/$MAE$ results  on the overall dataset. ART is the baseline for each algorithm. The highlighted row (ART) gives the outcome of the baselines.}
	\vspace{-2mm}
	\label{tab:performaceCompare}
	\begin{tabular}{|@{}c@{}|@{}c|@{}c|@{}c|@{}c|}
		\hline
		& \textbf{RF} & \textbf{SVR} & \textbf{NN} & \textbf{LR} \\
		\hline
		ALL & \textbf{0.560/0.282} &  0.472/0.324 & 0.499/0.310 & 0.413/0.338 \\
		\hline
		UC & 0.520/0.294 & 0.479/0.303 & 0.502/0.301 &  0.400/0.342  \\
		\Cline{1-5}
		\Thickvrulel{ART} & 0.078/0.439 & 0.021/0.452 & 0.016/0.459 & 0.020/0.458 \\ 
		\Cline{1-5}
		\texttt{rate} & 0.470/0.316 & 0.465/0.311 & 0.459/0.323 &  0.370/0.354   \\
		\hline
	\end{tabular}
	\vspace{-5pt}
\end{table}	

We report the performance for the four algorithms (i.e., Random Forest, Support Vector Regression, Neural Network, and Linear Regression) along the four sets of features (i.e., ALL, UC, ART, and \texttt{rate}), in the  \textit{global} setting in Table \ref{tab:performaceCompare}. We omit the outcome with  the \textit{local} setting because of the page limitation. But, the conclusion is very similar  to the one drawn in the \textit{global} setting.

\subsubsection{\textbf{User Factors Matter}} In Table \ref{tab:performaceCompare}, the combined use of all features (ALL) along with Random Forest achieves the best accuracy. We observe that the $R^2$ value for ART in each testing scenario is near zero, suggesting that the article features alone are not useful signals for predicting the comment volume an article will receive. 
The prediction models yield much better results when the commenting behavior from early users are taken into consideration (e.g., compare UC row to ART row). 
\emph{This proves that attempting to predict the eventual volume of user comments on the merits of a news article itself is a futile endeavour.} 
The reason is that a large fraction of the users post comments are triggered by other users' comments instead of the content of the news article itself \cite{singer09,RuizDMDMM11}. The article features cannot account for the user commenting dynamics. Thus, it is necessary to look into the early user commenting behavior after the publication of a news article to improve our ability to approximate the eventual comment volume of the article.

Since the difference among MAEs across the four feature spaces for a specific algorithm is not as clear as $R^2$, we use $R^2$ to discuss additional issues about the prediction performance in the subsequent sections. We use Random Forest in the remaining experiments.

\subsubsection{\textbf{Non-linearity}} Contrasting the performances of the linear algorithm (Linear Regression) and other nonlinear algorithms (Random Forest, Support Vector Regression, and Neural Network), the $R^2$ of Linear Regression is consistently worse no matter which feature space (except for ART) is considered. 
%This indicates that the task of predicting the eventual number of comment in a news article is not linear; hence, it is not an easy task.
This suggests that Linear Regression alone cannot solve the task of predicting the eventual comment number in a news article. 
We need to look into more complex models to improve accuracy.

\subsection{Dominant Feature Discovery}

\begin{table}[t]
	\small
	\centering
	\caption{$R^2$ results for feature ablation and selection for both global and local settings with machine learning algorithm Random Forest. 
		Acronyms: WSP: Washington Post, DM: Daily Mail, WSJ: Wall Street Journal, FN: Fox News, Gd: the Guardian, NYT: New York Times.
	}
	\vspace{-2mm}
	\label{tab:featureResult}
	\begin{tabular}{|c|c|c|c|c|@{}c@{}|}
		\hline
		\textbf{Model} & \textbf{ALL} & \textbf{UC} & \textit{\textbf{ALL $-$ UC}} & \textbf{\textit{rate}} & \tabincell{c}{\textbf{\textit{ALL $-$ \{rate\}}}}\\
		\hline
		WSP & 0.533 & 0.480 & 0.147 & 0.471 & 0.152 \\
		\hline
		DM & 0.541 & 0.498 & 0.170 & 0.477 & 0.193 \\
		\hline
		WSJ & 0.737 & 0.726 & 0.327 & 0.651 & 0.359 \\
		\hline
		FN & 0.449 & 0.408 & 0.142 & 0.378 & 0.134 \\
		\hline
		Gd & 0.468 & 0.428 & 0.168 & 0.416 & 0.170 \\
		\hline
		NYT & 0.631 & 0.612 & 0.213 & 0.484 & 0.280 \\
		\hline
		Overall & 0.560 & 0.520  & 0.185 & 0.470 & 0.209 \\
		\hline
	\end{tabular}
	\vspace{-5pt}
\end{table}

We perform feature ablation by removing one set of features at a time to understand the strengths of the feature families described in Table \ref{tab:features}. We report the outcome for both the global and local settings. Table \ref{tab:featureResult} summarizes the outcome of this study.
We observe that the decrease in $R^2$ is no more than 0.055 when we include only the comment features (compare the columns ALL and UC). The performance drops dramatically if we remove the comment features (see column \textit{ALL $-$ UC}). This is another supporting evidence on our account that signals gathered from early user comments largely influence the ability to predict comment volume.

We also study the importance of the individual features by applying the stepwise forward feature selection method \cite{guyonE03}. 
%\cite{kohaviJ1997,guyonE03}. 
This study shows that \texttt{rate} (from the group of comment features) is the most useful predictive variable in the prediction of comment volume. To further understand its importance, we redo the experiments with leaving out \texttt{rate}. The last column in Table \ref{tab:featureResult} displays the outcome. Compared with the results in the column ALL, there is a dramatic drop in $R^2$, from 0.560 to 0.209, in the global setting. The decrease ranges between 0.298 - 0.381 across the outlets. 
This further illustrates the importance of \texttt{rate} in the prediction task at hand.

The outcome of feature ablation and selection is consistent with the results in Table \ref{tab:performaceCompare}. We also draw the same conclusion from the other algorithms: \texttt{rate} is the dominant single feature in the prediction task.

\section{Rate Analysis}
\label{sec:rateStudy}

\begin{figure}[t]
	\centering
	%\vspace{-2mm}
	\includegraphics[width=0.9\linewidth, height=0.48\linewidth]{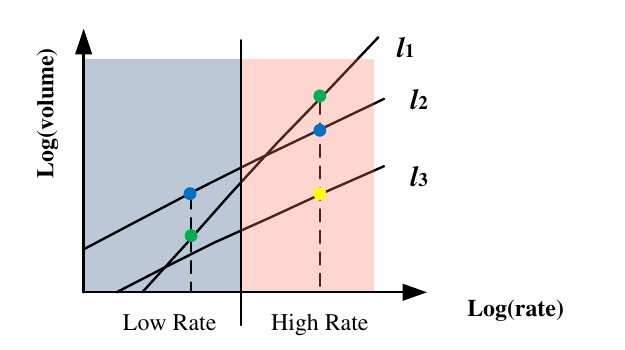}	
	\vspace{-4mm}
	\caption{Comparison of regression lines.}
	\label{fig:lineComparison}
	\vspace{-5pt}
\end{figure}

\begin{figure*}[!t]
	\centering
	\begin{subfigure}{0.5\textwidth}
		\includegraphics[width=0.85\linewidth, height=0.45\linewidth]{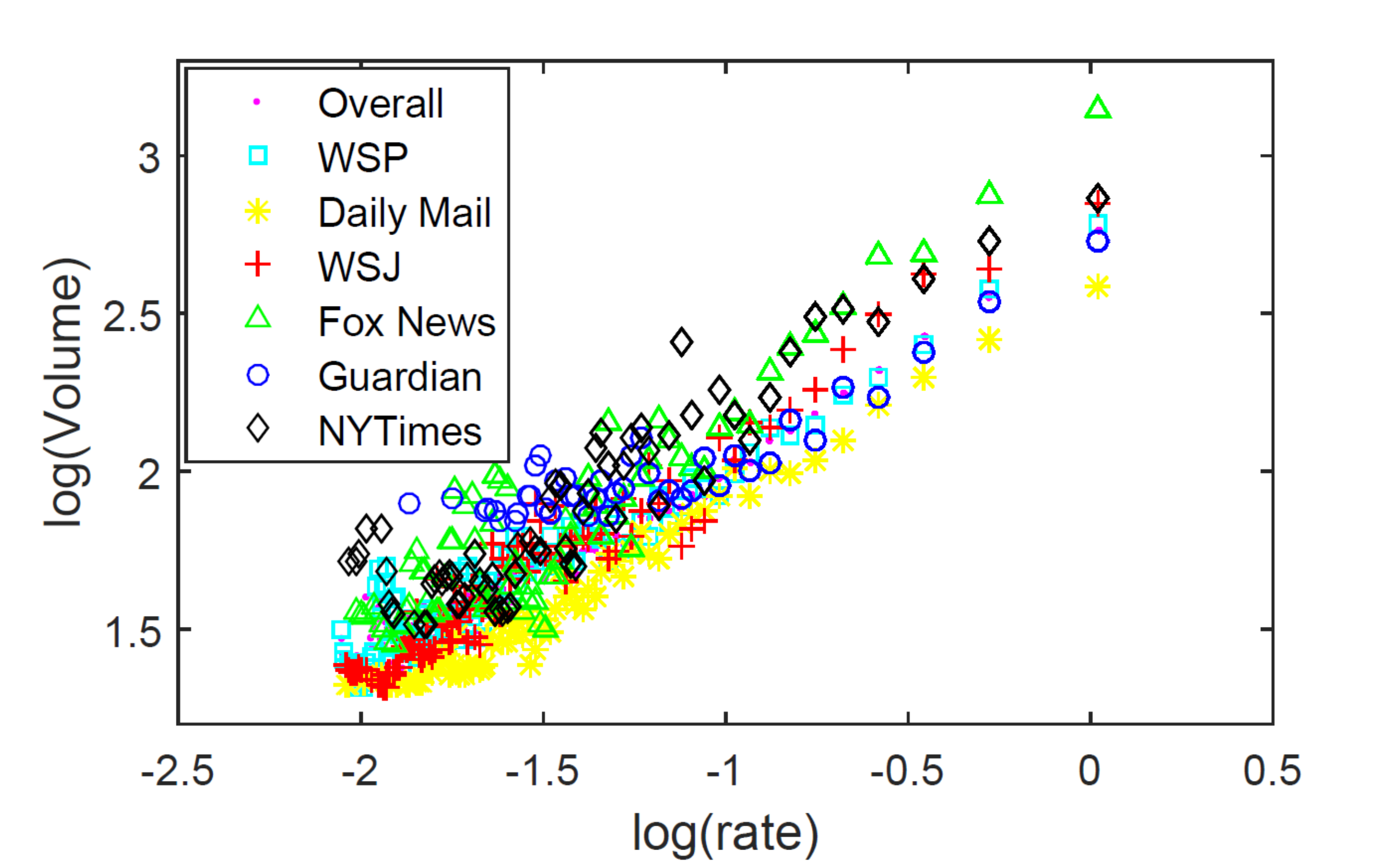}
		\vspace{-1mm}
		\caption{Rate and Volume in the log scale}	
		\label{fig:logRateVolumeOutlet}
	\end{subfigure}%
	\begin{subfigure}{0.5\textwidth}
		\includegraphics[width=0.85\linewidth, height=0.45\linewidth]{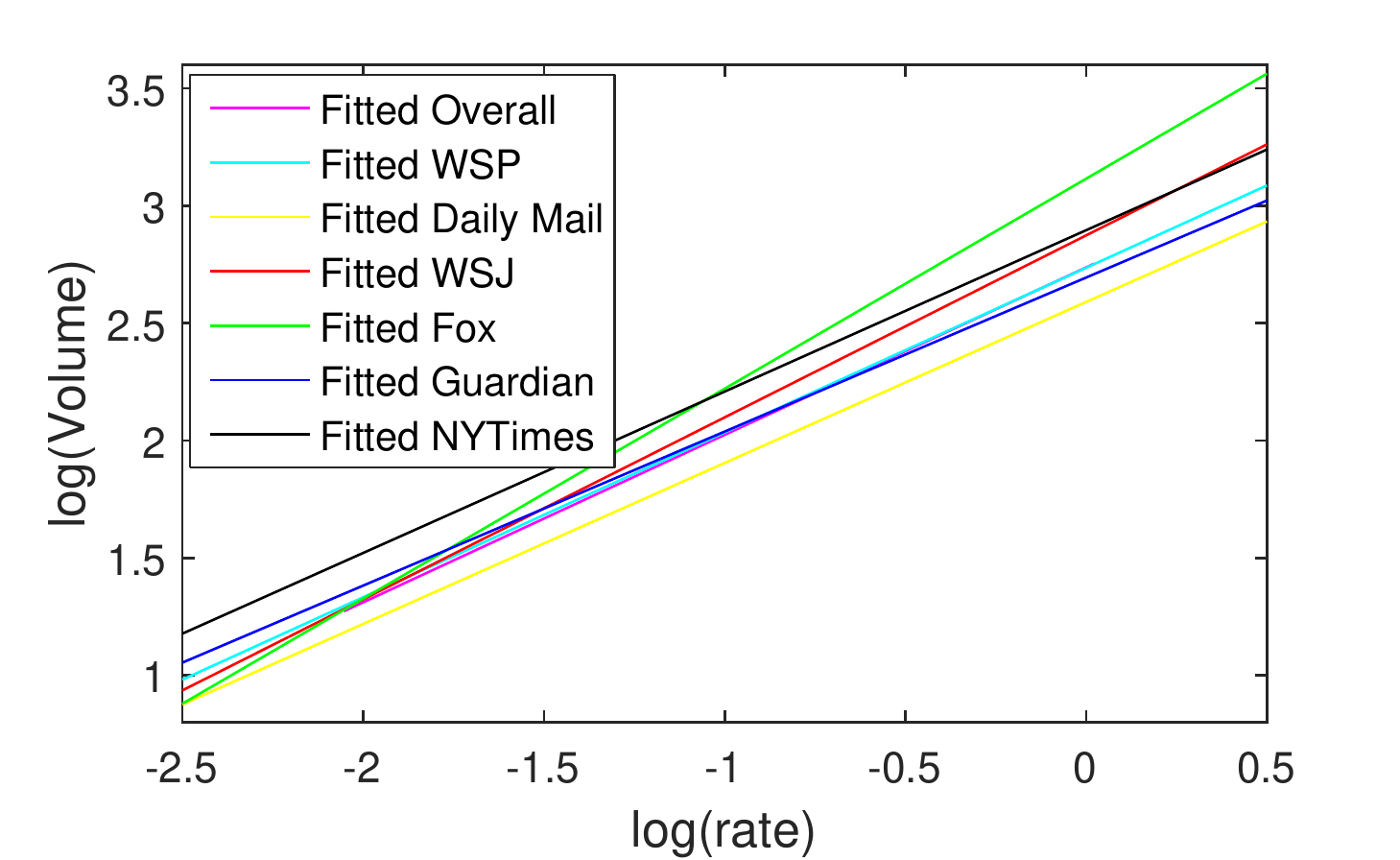}
		\vspace{-1mm}
		\caption{Fitted Lines}	
		\label{fig:fitlogRateVolumeOutlet}	
	\end{subfigure}
	\vspace{-3mm}
	\caption{The plot of the prediction model. If we plot the points based on the value of rate and logarithmic volume, points are too dense around the origin. Therefore, we draw the graph in the log scale for rate and volume.}
	\label{fig:predictionModel}
	\vspace{-5pt}
\end{figure*}

%From previous analysis, we know \textit{rate} is the key feature that contributes to the prediction of comment volume in articles.
In this section, we focus on the prediction models trained only with \texttt{rate}, and investigate their characteristics across news outlets and news categories. We use Random Forests in the experiments reported in this section. %Section \ref{sec:predictVolume} gives a comprehensive study of more than 30 features showing that  \texttt{rate} is the dominant feature that contributes over 90\% to the  prediction accuracy in most cases.

%We explore a set of features in the prediction of comment volume in a news article and find that ``the arrival rate of \emph{early} comments" (\texttt{rate}), defined as the number of comments per minute in the early commenting stage, is the dominant feature that contributes to the prediction task (we prove it in Section \ref{sec:predictVolume}, please believe our conclusion here).

\subsection{Study of Rate across Outlets}

We build rate models for both the global and local settings with Random Forest, considering the observed \texttt{rate}'s among the first $\alpha$ = 10 user comments, and use them to predict the eventual comment volume of a news article.
%Taking rate back to the model, we then get the predicted number. Therefore, we are able to know how the model looks like based on the value of rate and predicted result.
We also repeat the studies of rate models built on other values of $\alpha$ (the results are provided in Appendix), and the results are consistent with the 10 comment threshold.

\subsubsection{\textbf{Rate Modeling}}
\label{subsec:rateModel}
%The appearance of prediction model in each dataset is shown in Figure \ref{fig:rateVolumeOutlet}. The best fit type of the models in general is power function as the curves in the graph indicate. According to the nature of power function, if we plot the value of rate and predicted volume in logarithmic scale over all points in each dataset, they should be in linear relationship and best fitted by a linear regression line (displayed by a slope and intercept).

Figure \ref{fig:logRateVolumeOutlet} shows the shape of the prediction models. Points in each dataset are fitted by a linear regression line, as shown in Figure \ref{fig:fitlogRateVolumeOutlet}.
We use the cartoon example in Figure \ref{fig:lineComparison} to describe the chief points we seek to convey in this study about the rate models. There are three regression lines in Figure \ref{fig:lineComparison}. \textit{l1} is the fitted regression line of points in outlet 1, \textit{l2} is for outlet 2, and \textit{l3} for outlet 3.
We distinguish two interesting cases: (1) the regression lines cross each other, as in \textit{l1} and \textit{l2}; and (2) the regression lines are parallel, as in \textit{l2} and \textit{l3}.
%Suppose \textit{l1} is steeper than \textit{l2}, and the intercept of \textit{l2} is larger than \textit{l3}.

Consider the lines \textit{l1} and \textit{l2}. The rate area can be split into two parts: low rate and high rate, based on their intersection. If we carefully compared the points on the two lines, we gather that the user commenting behaviors in outlets 1 and 2 vary across areas. In the low rate area, users in both outlets show less interest at the beginning, reflected by the small values of rate, but the users in outlet 2 keep commenting more than those in outlet 1 as indicated by the larger eventual comment volume. However, in the high rate area, articles in outlet 1 attract more commenting activity than those in outlet 2, even though the commenting activity early on is the same.

We can draw another useful observation by studying the lines \textit{l1} and \textit{l2}:
%of the user behavior in outlet 1 and 2 is that
the same rate fluctuation leads to different variation in comment volume. Since the slope of \textit{l1} is larger than that of \textit{l2}, \textit{l1} will grow faster. Therefore, we conclude that the comment volume in outlet 1 is more sensitive to the rate in the early commenting stream than the comment volume in outlet 2.

Consider the parallel lines \textit{l2} and \textit{l3}. This scenario suggests that the same initial rate leads to different commenting volume in outlets 2 and 3. More precisely, the comment volume of a news article from outlet 2 is larger than that of a news article from outlet 3.

\subsubsection{\textbf{Comparison across Outlets}}
\label{sec:ComparisonAcrossOutlets}

\begin{table}[t]
	\small
	\centering
	\caption{Statistics of the prediction model trained by \textit{rate}. The values in column Slope (Intercept) Interval are the lower and upper confidence limits for 95\% confidence intervals of Slope (Intercept). Column MoPV = Mean of Predicted Volume in the log scale. We reuse the acronyms for news outlets in Table \ref{tab:featureResult}.}
	\label{tab:p1p2outlet}
	\vspace{-2mm}
	\begin{tabular}{|@{}c@{}|c|@{}c@{}|@{}c|@{}c|@{}c@{}|}
		\hline
		\textbf{Model} & \textbf{Slope} & \textbf{Intercept} & \tabincell{c}{\textbf{Slope}\\\textbf{Interval}} & \tabincell{c}{\textbf{Intercept}\\\textbf{Interval}} & \textbf{MoPV} \\		
		\hline
		WSP & 0.758 & 2.740 & [0.755, 0.762] & [2.766, 2.768] & 2.163 \\
		\hline
		DM & 0.703 & 2.606 & [0.701, 0.705] & [2.604, 2.608] & 2.131\\
		\hline
		WSJ & 0.841 & 2.885 & [0.832, 0.849] & [2.875, 2.895] & 2.111 \\
		\hline
		FN & \textbf{0.963} & \textbf{3.201} & [0.953, 0.972] & [3.194, 3.209] &  \textbf{2.577} \\
		\hline
		Gd & 0.656 & 2.728 & [0.649, 0.663] & [2.723, 2.732] & 2.396 \\
		\hline
		NYT & 0.707 & 2.935 & [0.695, 0.719] & [2.924, 2.945] & 2.454 \\
		\hline
		Overall & 0.777 & 2.767 & [0.776, 0.778] & [2.766, 2.768] &  2.224\\
		\hline		
	\end{tabular}
	\vspace{-5pt}
\end{table}	

\begin{comment}
\begin{table}[t]
	\small
	\centering
	\caption{Statistics of the prediction model trained by \textit{rate}. MoPV = Mean of Predicted Volume in the log scale.}
	\label{tab:p1p2outlet}
	\vspace{-4mm}
	\begin{tabular}{|c|c|c|c|}
		\hline
		\textbf{Model} & \textbf{Slope} & \textbf{Intercept} & \textbf{MoPV} \\		
		\hline
		Washington Post & 0.701 & 2.737 & 2.163 \\
		\hline
		Daily Mail & 0.683 & 2.589 & 2.131\\
		\hline
		Wall Street Journal & 0.779 & 2.880 & 2.111 \\
		\hline
		Fox News & \textbf{0.894} & \textbf{3.115} & \textbf{2.577} \\
		\hline
		the Guardian & 0.656 & 2.694 & 2.396 \\
		\hline
		New York Times & 0.686 & 2.895 & 2.454 \\
		\hline
		Overall & 0.713 & 2.737  & 2.224\\
		\hline		
	\end{tabular}
	\vspace{-10pt}
\end{table}	
\end{comment}

We plot the shape of rate models in Figure \ref{fig:predictionModel} and provide the slopes and intercepts of regression lines in Table \ref{tab:p1p2outlet}. We provide the lower and upper confidence limits for the 95\% confidence intervals of slope and intercept of regression lines for each dataset. The bounds of slopes and intercepts show that their observed differences are due to actual differences between outlets rather than random chance. We also calculate the mean of the predicted logarithm of the comment volume for each dataset (column MoPV in Table \ref{tab:p1p2outlet}). We present the results in the light of the discussion in the previous section. The plot shows that the users at Fox News are more active than those at the other news outlets--- the logarithm of volume for Fox News is 2.577, which is close to the true value 2.47 in Table \ref{tab:summary}.

The regression line of Fox News (Figure \ref{fig:fitlogRateVolumeOutlet}) behaves as line \textit{l1} in Figure \ref{fig:lineComparison}. 
Its high slope suggests that the total comment volume is very sensitive to the \texttt{rate} of the initial comments at this outlet. If the rate is small, the discussion dies out quickly. If the rate is large, the readers become very engaged. 
%It seems that readers at Fox News are passionate and reactive. Readers at Wall Street Journal are ranked as the second most reactive, while readers at the Guardian are ranked as the least reactive (referring to its slopes in Table \ref{tab:p1p2outlet}).
The comment volume of an article from the Guardian is the least sensitive to the \texttt{rate} of early user comments.
Comparing the commenting activity at these outlets, the commenting activity at New York Times has the longest attention, because the commenting persists longer and this results in higher volume.  
%One explanation is that this is a property of its readership (readers think about articles longer), but it may also be an indicator of some latent variables (e.g., the quality or temporal relevance decay of articles), so they stay relevant longer in New York Times than in other outlets.
We think that the observed behavior might be caused by factors such as the quality or temporal relevance decay of articles, which are difficult to extract from the article content.
New York Times behaves as \textit{l2} and Daily Mail as \textit{l3} in Figure \ref{fig:lineComparison}.
The commenting attention at Daily Mail seems to be lower. 
Interestingly, Wikipedia calls Daily News ``middle-market tabloid,"  ``that attempts to cater to readers who want some entertainment from their newspaper," so it might indicate a more fleeting character of their articles compared to other outlets.

%We note that the regression line of Fox News (Figure \ref{fig:fitlogRateVolumeOutlet}) is almost persistently above the others, behaving as line \textit{l1} from Figure \ref{fig:lineComparison}. This is also indicated by its intercept in Table \ref{tab:p1p2outlet}, which is the largest. It suggests that the users at Fox News maintain longer a higher commenting activity than the users at the other news outlets. The high slope of the regression line of Fox News suggests that the predicted volume is very sensitive to the initial comments, which means that a small change in early rate will lead to a larger variation in eventual volume. The regression lines of Daily Mail and New York Times are parallel, while the intercept of the latter is larger than that of the former. This indicates that the user commenting lifespan in New York Times is longer than that in Daily Mail. This is consistent with the conclusion derived when comparing the mean values of the volumes at the two news outlets.
%which is also the case implied by the mean of comment volume in Table \ref{tab:summary}.

\subsubsection{Article Examples for Rate Models}

\begin{table}[t]
	\small
	\centering
	\caption{Article examples for outlets with the regression lines crossing each other (Fox News vs the Guardian) and in parallel (Daily Mail vs New York Times).}
	\label{tab:articleExampleOfRate}
	\vspace{-2mm}
	\begin{tabular}{|c|c|c|c|}
		\hline
		& & \textbf{Fox News}  & \textbf{the Guardian} \\
		\hline
		\multirow{3}{*}{Pair 1} & Article & $FN_1$   & $GD_1$  \\\cline{2-4}
		& Rate & 0.769 & 0.769 \\\cline{2-4}
		& $N_A$ & 2,768 & 705 \\\cline{2-4}
		\hline
		\multirow{3}{*}{Pair 2} & Article & $FN_2$   & $GD_2$  \\\cline{2-4}
		& Rate & 0.092 & 0.092 \\\cline{2-4}
		& $N_A$ & 30 & 117 \\\cline{2-4}
		\hline
		\hline
		& & \textbf{Daily Mail}  & \textbf{New York Times} \\
		\hline
		\multirow{3}{*}{Pair 3} & Article & $DM_1$  & $NYT_1$ \\\cline{2-4}
		& Rate & 0.667 & 0.667\\\cline{2-4}
		& $N_A$ & 227  & 407 \\\cline{2-4}
		\hline
		\multirow{3}{*}{Pair 4} & Article & $DM_2$  & $NYT_2$  \\\cline{2-4}
		& Rate & 0.07 & 0.07 \\\cline{2-4}
		& $N_A$ & 50 & 108 \\\cline{2-4}
		\hline
	\end{tabular}
	\vspace{-5pt}
\end{table}	

We provide a few concrete examples to illustrate the rate-to-volume behavior across pair of news outlets in this section.
We illustrate two scenarios: (i) pairs of news article for outlets whose regression lines cross each other (e.g., Fox News vs the Guardian) and (ii) pairs of news article for outlets whose regression lines are parallel (e.g., Daily Mail vs New York Times) in Table \ref{tab:articleExampleOfRate}. For each of (i) and (ii) we showcase pairs of news articles from both the high and low rate zones.

%According to the previous analysis, the study of the rate models for the crossed situation consider the rate area (high or low). 

\textbf{Crossing regression lines.} We have two pairs of news articles ($FN_1$, $GD_1$)\footnote{\tiny{\url{www.foxnews.com/politics/2016/01/12/in-gop-response-haley-pans-obama-presidency-makes-case-for-new-direction.html}}}$^,$\footnote{\tiny{\url{theguardian.com/politics/2015/nov/26/labour-whip-email-vote-against-syria-airstrikes}}}  and ($FN_2$, $GD_2$)\footnote{\tiny{\url{www.foxnews.com/opinion/2015/12/01/in-paris-obama-worships-at-altar-europes-real-religion-climate-change.html}}}$^,$\footnote{\tiny{\url{www.theguardian.com/science/2015/oct/28/us-approval-for-drug-that-turns-herpes-virus-against-cancer}}} from Fox News and the Guardian, respectively.
%: $A_1^{FN}$ vs $A_1^{GD}$ 
The rates of ($FN_1$, $GD_1$) are both high at 0.769, while the rates of ($FN_2$, $GD_2$) are low at 0.092. According to our analysis of the rate-to-volume behavior, since the rates of $FN_1$ and $GD_1$ are located in the high rate area and the slope of the regression line for Fox News is larger than that of the Guardian, we expect $N_A$, the eventual number of comments, of $FN_1$ to be larger than that of $GD_1$: $FN_1$ receives more comments than $GD_1$, 2,768 versus 705 (in a week). See Pair 1 in Table \ref{tab:articleExampleOfRate}. The effect is reversed for $FN_2$ and $GD_2$:  $FN_2$ receives fewer comments than $GD_2$, 30 versus 117. See Pair 2 in Table \ref{tab:articleExampleOfRate}.

%Based on the rate models of Fox News and the Guardian, we know the prediction of $N_A$ for $A_1^{FN}$ would be larger than that of $A_1^{GD}$, because these two articles are located in the high rate area and the slope of the regression line for Fox News is larger than the Guardian. Since $A_2^{FN}$ and $A_2^{GD}$ are in the low rate area, $N_A$ of $A_2^{GD}$ will be predicted larger than that of $A_2^{FN}$. To see whether the prediction from the rate models is correct, we have the true value of $N_A$ for each article in Table \ref{tab:articleExampleOfRate}. It shows $A_1^{FN}$ receives more comments than $A_1^{GD}$ while comment volume in $A_2^{FN}$ is less than $A_2^{GD}$, which supports the conclusion we have from rate models.

\textbf{Parallel regression lines.} For the parallel case (Daily Mail vs New York Times), we also provide two pairs of news articles: ($DM_1$, $NYT_1$)\footnote{\tiny{\url{www.dailymail.co.uk/news/article-3296561/Syrian-anti-ISIS-activist-blogged-terrible-conditions-Raqqa-decapitated-Turkey-alongside-beheaded-corpse-friend.html}}}$^,$\footnote{\tiny{\url{www.nytimes.com/2015/11/12/us/politics/republicans-ted-cruz-marco-rubio.html}}} and ($DM_2$, $NYT_2$)\footnote{\tiny{\url{www.dailymail.co.uk/sciencetech/article-3311075/Anomalies-thermal-scanning-Egypt-pyramids.html}}}$^,$\footnote{\tiny{\url{www.nytimes.com/2016/02/09/sports/basketball/knicks-fire-derek-fisher-as-coach.html}}}. The rates of ($DM_1$, $NYT_1$) are both high at 0.667, while the rates of ($DM_2$, $NYT_2$) are low at 0.07. Since the regression lines of these two outlets are parallel, we do not expect a reversal as in the previous case. Hence, we expect both the $N_A$ of $NYT_1$ to be larger than that of $DM_1$ (407 versus 227, Pair 3 in the table), and the $N_A$ of $NYT_2$ to be larger than that of $DM_2$ (108 versus 50, Pair 4 in the table).

%Their rate models suggest that the news articles from New York Times will always receive more comments than Daily Mail once they have the same arrival rate of early comments, which is consistent with the true comment volume in Table \ref{tab:articleExampleOfRate} (see $N_A$ of Pair 3 and Pair 4).

\subsection{Study of Rate across Categories}

\begin{comment}
\begin{table}[t]
\centering
\caption{Statistics of the datasets and the models trained by \textit{rate} in category domains. MoPV is the mean of the predicted logarithmic volume.}
\label{tab:r2p1p2category1}
\vspace{-3mm}
\begin{tabular}{|@{}c@{}|@{}c@{}|c|c|@{}c@{}|c|}
\hline
\textbf{Category} & \tabincell{c}{\textbf{\# of} \\ \textbf{Articles}} & \textbf{$R^2$} & \textbf{Slope} & \textbf{Intercept} & \textbf{MoPV}  \\
\hline		
Politics & 8,491 & 0.442 & 0.7306 & 2.8351 & 2.321 \\
\hline
US & 6,202 & 0.457 & 0.6657 & 2.7324 & 2.259 \\
\hline
World & 2,548 & 0.490 & 0.7015 & 2.7515 & 2.195 \\
\hline
Sports & 1,839 & 0.484 & 0.5687 & 2.4579 & 1.929 \\
\hline
Entertainment & 1,828 & 0.535 & 0.6438 & 2.5640 & 2.089 \\
\hline
Technology & 469 & 0.528 & 0.6314 & 2.5784 & 1.977 \\
\hline
Business & 342 & 0.472 & 0.6695 & 2.6592 & 2.078 \\
\hline		
Science & 235 & 0.468 & 0.6522 & 2.6410 & 2.035 \\
\hline
Health & 156 & 0.493 & 0.5565 & 2.4912 & 1.872 \\
\hline
\end{tabular}
\vspace{-3mm}
\end{table}	
\end{comment}

We now study the performance and characteristics of rate models by news category. We show that rate model behaves quite differently across the major news categories.
The present results are from the rate models built on the first 10 user comments, but we have consistent observations from other values of $\alpha$.

\subsubsection{\textbf{Categorizing Articles}}
To analyze \texttt{rate} in different news categories, we need to assign each article to its corresponding categories first. We observe that news outlets assign category labels to their articles. Our initial idea was to make use of these category labels. However, we soon noticed that labels are not consistent across news outlets. For example, the categories such as ``U.S. Showbiz" in Daily Mail, ``Local" and ``National" in Washington Post, 
%``Economy" in Wall Street Journal, 
``Soccer" in the Guardian, and ``Magazine" in New York Times are unique to these outlets. It is difficult to confidently align these categories over time in general, because news outlets periodically reorganize their news categories.
%not reasonable to keep all category labels provided in these news outlets.
We thus resort to the category labels in Google News, which are more stable over a longer period of time. We set the set of labels $C$ =  \{``Politics", ``US", ``World", ``Sports", ``Entertainment", ``Technology", ``Business", ``Science", ``Health"\} in our analysis. We add ``Politics" because it is a category that appears uniformly in all news outlets.
We encounter many articles which are not explicitly assigned to any of these categories. We use their topics to determine their categories. We first categorize the topics of an article and then propagate the category labels to the article. We describe the process below.

\textbf{Categorizing Topics.} We follow the method of categorizing topics proposed in \cite{zhaoJWHLYL11}. For a topic $t$, its probability of belonging to category $q \in C$ is
$$  p(q|t) = \frac{p(q,t)}{p(t)} = \frac{|D_{q,t}|}{|D_t|} \propto |D_{q,t}|, $$
where $D_t$ is the union of articles whose topic is $t$, $D_{q,t}$ denotes the subset of articles in $D_t$ that are labeled
with category $q$. For a specific topic $t$, $D_t$ is constant. We select the top three categories for $t$ if it belongs to over three ones. For example, the topic ``Donald Trump" is assigned to the categories ``US," ``World," and ``Politics."

\textbf{Assigning News Categories to News Articles.}
Our crawler extracts topic information for each news article. Once we determine the news categories $C_t$ of a topic $t$, we place each article on topic $t$ in each of the categories in $C_t$.
%However, if the topic's category does not fall into set $C$, we discard the corresponding articles under this topic. In this way, the category of each article is in the set $C$.

\subsubsection{\textbf{Comparison across Categories}}

\begin{comment}
\begin{table}[t]
	\small
	\centering
	\caption{Statistics of the datasets and the global models trained by \textit{rate} in category domains. }
	\label{tab:r2p1p2category}
	\vspace{-4mm}
	\begin{tabular}{|c|c|c|c|c|c|}
		\hline
		\textbf{Category} & \tabincell{c}{\textbf{\# of} \\ \textbf{Articles}} & \textbf{$R^2$} & \textbf{Slope} & \textbf{Intercept}  \\
		\hline		
		Politics & 8,491 & 0.442 & 0.731 & 2.835 \\
		\hline
		US & 6,202 & 0.457 & 0.666 & 2.732 \\
		\hline
		World & 2,548 & 0.490 & 0.702 & 2.752 \\
		\hline
		Sports & 1,839 & 0.484 & 0.569 & 2.458  \\
		\hline
		Entertainment & 1,828 & 0.535 & 0.644 & 2.564 \\
		\hline
		Technology & 469 & 0.528 & 0.631 & 2.579  \\
		\hline
		Business & 342 & 0.472 & 0.669 & 2.659  \\
		\hline		
		Science & 235 & 0.468 & 0.652 & 2.641  \\
		\hline
		Health & 156 & 0.493 & 0.556 & 2.491  \\
		\hline
	\end{tabular}
	\vspace{-10pt}
\end{table}	
\end{comment}

\begin{table}[t]
	\small
	\centering
	\caption{Statistics of the datasets and the global models trained by \textit{rate} in category domains. The values in column Slope (Intercept) shows the slope (intercept) of the \textit{rate} model together with its lower and upper confidence limits for 95\% confidence intervals.}
	\label{tab:r2p1p2category}
	\vspace{-2mm}
	\begin{tabular}{|@{}c@{}|@{}c@{}|c|@{}c|@{}c@{}|}
		\hline
		\textbf{Category} & \tabincell{c}{\textbf{\# of} \\ \textbf{Articles}} & \textbf{$R^2$} & \tabincell{c}{\textbf{Slope} \\ (Interval)} & \tabincell{c}{\textbf{Intercept} \\ (Interval)}  \\
		\hline		
		Politics & 8,491 & 0.442 & \tabincell{c}{0.729\\([0.727, 0.730])} & \tabincell{c}{2.818\\([2.816, 2.819])} \\
		\hline
		US & 6,202 & 0.457 & \tabincell{c}{0.667\\([0.665, 0.668])} & \tabincell{c}{2.719\\([2.717, 2.720])} \\
		\hline
		World & 2,548 & 0.490 & \tabincell{c}{0.700\\([0.696, 0.704])} & \tabincell{c}{2.735\\([2.732, 2.739])} \\
		\hline
		Sports & 1,839 & 0.484 & \tabincell{c}{0.570\\([0.566, 0.575])} & \tabincell{c}{2.448\\([2.444, 2.453])}  \\
		\hline
		Entertainment & 1,828 & 0.535 & \tabincell{c}{0.645\\([0.640, 0.650])} & \tabincell{c}{2.549\\([2.545, 2.554])} \\
		\hline
		Technology & 469 & 0.528 & \tabincell{c}{0.633\\([0.620, 0.647])} & \tabincell{c}{2.567\\([2.553, 2.581])}  \\
		\hline
		Business & 342 & 0.472 & \tabincell{c}{0.664\\([0.649, 0.679])} & \tabincell{c}{2.639\\([2.623, 2.654])}  \\
		\hline		
		Science & 235 & 0.468 & \tabincell{c}{0.652\\([0.626, 0.678])} & \tabincell{c}{2.626\\([2.599, 2.653])}  \\
		\hline
		Health & 156 & 0.493 & \tabincell{c}{0.562\\([0.525, 0.600])} & \tabincell{c}{2.485\\([2.440, 2.530])}  \\
		\hline
	\end{tabular}
	\vspace{-5pt}
\end{table}

We first partition the overall dataset across the news categories $C$, and then train a rate model in each category. The analysis follows the steps of the analysis across news outlets. Table \ref{tab:r2p1p2category} gives the article count, performance ($R^2$), and the characteristics of the rate models per news category.

%The large number of article size in ``Politics" point out people's high engagement there, which supports the statement in \cite{Jeff16,ekmanA12} that there is a growing interest in political participation. The large quantity of category ``Politics" and ``US" is attributed to the 2016 presidential election in America, which produces hot topics as ``Donald Trump", ``Hillary Clinton", and so on. In spite of the abundant instances, the performance of the rate model is the worst in these two categories (about 0.45).

The first observation is that $R^2$ differs very little across the news categories, hovering about 0.5. This is in stark contrast to the observations made about $R^2$ in news outlets.
%, where $R^2$ varies between .378 to .651 (see column \textit{rate} in Table \ref{tab:featureResult}).
%The predicted comment number is less than the truth in average, as what we observed in outlets. But it captures the specialties of categories as well, such as the largest volume in ``Politics" and the smallest in ``Health".
Comparing the linear regression in each category, ``Politics" has the highest slope, indicating that the predicted comment volume for articles in this category is more sensitive to the early rate than for articles in other news categories.
%We also observe that people are reactive to news articles from categories ``US" and ``World" (as indicated by their large slopes and intercepts).
``Health" has the lowest comment volume and the prediction for its articles is less sensitive to the rate of early comments.

\subsection{Interplay between Outlets and Categories}

\begin{table}[t]
	\small
	\centering
	\caption{The distribution of article count by outlet and category. We reuse the acronyms for news outlets in Table \ref{tab:featureResult}.
		%WSP: Washington Post, DM: Daily Mail, WSJ: Wall Street Journal, FN: Fox News, Gd: Guardian, NYT: New York Times.
	}
	\label{tab:sizeOutletCategory}
	\vspace{-2mm}
	\begin{tabular}{|@{}c@{}|c|c|c|c|c|c|}
		\hline
		& \textbf{WSP} & \textbf{DM} & \textbf{WSJ} & \textbf{FN} & \textbf{Gd} & \textbf{NYT} \\	
		\hline		
		Politics & 3,374 & 2,077 & 1,212 & 940 & 408 & 480 \\
		\hline
		US & 2,257 & 1,548 & 966 & 283 & 655 & 493 \\
		\hline
		World & 635 & 878 & 362 & 188 & 351 & 134 \\
		\hline
		Sports & 403 & 1,060 & 112 & 62 & 154 & 102 \\
		\hline
		Entertainment & 199 & 1,242 & 53 & 80 & 215 & 39 \\
		\hline
		Technology & 80 & 195 & 88 & 42 & 48 & 16  \\
		\hline
		Business & 80 & 105 & 89 & 8 & 41 & 19 \\
		\hline		
		Science & 33 & 100 & 33 & 31 & 29 & 9 \\
		\hline
		Health & 33 & 72 & 18 & 25 & 4 & 4 \\
		\hline
	\end{tabular}
	\vspace{-5pt}
\end{table}	

Given the different performance of the rate models across outlets and news categories, a natural follow up question is whether there is some mutual effect between outlets and categories over rate. For this study, we summarize the distribution of article sizes among outlets and categories in Table \ref{tab:sizeOutletCategory}. Then, we repeat the prediction analysis for rate models. Table \ref{tab:r2p1p2OutletCategory} displays the results per outlet and news categories.

Comparing Table \ref{tab:r2p1p2OutletCategory} (local rate models) to Table \ref{tab:r2p1p2category} (global rate models), we observe that the local models at Washington Post achieve similar $R^2$ as the global one in ``Politics" and ``US." For ``Sports," local models at Daily Mail perform similarly to the global one in Table \ref{tab:r2p1p2category}.
This is because either Washington Post or Daily Mail dominates the article size in these categories (see Table \ref{tab:sizeOutletCategory}).
%Looking at the article sizes in Table \ref{tab:sizeOutletCategory}, Washington Post and Daily Mail provide most of the data in each category. In Table \ref{tab:r2p1p2OutletCategory}, we notice that the global rate model achieves similar performance as the local ones in Washington Post and the categories ``Politics," ``US," and ``World,"  (Table \ref{tab:r2p1p2category}). The local models in Daily Mail have similar $R^2$ as the global one in ``Sports" and ``Entertainment" (Table \ref{tab:r2p1p2category}). This is because Daily Mail dominates the article size in these categories. The rate model in Wall Street Journal performs the best, especially in ``Politics" and ``US".

We notice that commenting activity in the ``Politics" articles at Fox News is quite distinct from others: by far the largest slope and intercept.
%(much higher than for ``US" and ``World"). 
In other outlets, ``Politics", ``US," and ``World" are comparable. This indicates that the total comment volume of an article from ``Politics" is very sensitive to the \texttt{rate} of the initial comments at Fox News.

User commenting preference per category within each outlet reveals additional properties about news outlets (Table \ref{tab:sizeOutletCategory}).
%On one hand, most of the articles in Fox News are talking about politics.
%We think that the attention of political events there is due to its ideological slant and the significant effect in Presidential elections, as mentioned in \cite{dellavignaK07,iyengarH09}.
%On the other hand, articles from Daily Mail display attention in sports and entertainment as well.
In Table \ref{tab:r2p1p2OutletCategory}, we notice that the value of intercept in ``Politics" is the largest across all categories and outlets, except for Daily Mail where ``World" has the largest intercept. Besides, the highest slope appears three times in ``Politics", twice in ``World", and once in ``US". Considering that most of the ``US" and ``World" articles are related to politics, it is reasonable to conclude that the comment volume is \emph{sensitive to rate} (suggested by high slope) and \emph{higher} (reflected by large intercept) in political area at most outlets.

%From previous analysis, we know that the influence of rate is significant in political domain. It is intuitive to ask how the \texttt{rate} model performs in each outlet without considering politics articles. Removing articles in ``Politics", we find that $R^2$ improves obviously in the global model, Daily Mail, and New York Times, and stays unchanged in Washington Post and Guardian, while drops in Wall Street Journal and Fox News. This observation implies that the effect of the sensitive dependency to \texttt{rate} in political domain varies across outlets.
%This implies that the sensitive dependency to rate in political domain is a double-edged sword. It may make the prediction better (as in Wall Street Journal and Fox News) or worse (as in the global model, Daily Mail and New York Times).

\section{Conclusion}
\label{sec:conclusion}

In this paper, we study the problem of predicting the total number of user comments a news article will receive.
We compile and analyze a large set of features, which we group by topic, article, user comment, news factor, and miscellaneous.
Our main insight is that the early dynamics of user comments contribute the most to an accurate prediction, while news article specific factors have surprisingly little influence. 
Furthermore, we show that the early arrival rate of comments is the best indicator of the eventual number of comments. We conduct an in-depth analysis of this feature across several dimensions, such as news outlets and news article categories.
We show that the prediction of comment volume is very sensitive to the early user commenting activity in some news outlets (e.g., Fox News) and categories (e.g., Politics).

%Given the dominant power of the predictive variable \texttt{rate}, we conduct an in depth analysis of its properties by looking at its behavior across news outlets and categories. We show that the prediction of comment volume is very sensitive to the early user commenting activity in some news outlets (e.g., Fox News) and categories (e.g., Politics). To prove the key role of \texttt{rate} in the prediction task, we study a large set of features, which we group by topic, article, comment, news factor, and miscellaneous. The feature set includes both old and novel (11 of them) features. We report their combined performance on the prediction task. Our experiments show conclusively that (1) one cannot rely on article features alone and ignore user factors when aiming to predict the comment volume, and (2) among all the features, the arrival rate of the early comments is the variable with the strongest predictive power. To our knowledge, these are novel discoveries. 

We believe that our findings shed new light on the unique characteristics of readership community compared to other online communities, e.g., those at Twitter or Facebook. This is particularly emphasized by the strong role of early user posting activity on the eventual comment volume for a news article. This has important implications in social media and user behavioral response process understanding, which are key components of the social-news media ecosystem. We believe that this insight is also of value to news analytics, which may lead to better understanding of user participation motives and engagement in commenting news items online. 
%The dynamics explored here have the potential to be leveraged and serve useful in predictive analytics and gauging social response to future events which have implications for social science.

%There is still plenty of room for improvement in the task of predicting the user comment volume of a news article. We plan to study the remaining news factor dimensions, such as unexpectedness and controversy, which require sophisticated NLP tools to mine signals specific to these classes from text, news article, and comment. 

\begin{table}[t]
	\small
	\centering
	\caption{The results of the local models trained by \textit{rate} among outlets and categories. The three values in each cell are $R^2$, then slope, and intercept. Some categories are removed because of insufficient articles. %We reuse the acronyms for news outlets in Table \ref{tab:featureResult}.
	}
	\label{tab:r2p1p2OutletCategory}
	\vspace{-2mm}
	\begin{tabular}{|c|c|c|c|c|c|c|}
		\hline
		& \textbf{WSP} & \textbf{DM} & \textbf{WSJ} & \textbf{FN} & \textbf{Gd} & \textbf{NYT} \\	
		\hline		
		Politics & \tabincell{c}{0.450\\0.701\\2.772} & \tabincell{c}{0.366\\0.671\\2.675} & \tabincell{c}{\textbf{0.657}\\0.783\\2.934} & \tabincell{c}{0.352\\\textbf{0.891}\\\textbf{3.202}} & \tabincell{c}{0.396\\0.657\\2.775} & \tabincell{c}{0.400\\0.647\\2.953} \\
		\hline
		US & \tabincell{c}{0.438\\0.652\\2.692} & \tabincell{c}{0.393\\0.652\\2.612} & \tabincell{c}{\textbf{0.628}\\0.772\\2.921} & \tabincell{c}{0.324\\0.690\\2.932} & \tabincell{c}{0.432\\0.744\\2.764} & \tabincell{c}{0.372\\0.628\\2.925} \\
		\hline
		World & \tabincell{c}{0.490\\0.714\\2.765} & \tabincell{c}{0.361\\0.707\\2.744} & \tabincell{c}{\textbf{0.587}\\0.706\\2.722} & \tabincell{c}{0.236\\0.663\\2.772} & \tabincell{c}{0.458\\0.634\\2.714} & \tabincell{c}{0.380\\0.640\\2.781} \\
		\hline
		Sports & \tabincell{c}{0.389\\0.566\\2.478} & \tabincell{c}{0.489\\0.588 \\2.426} & \tabincell{c}{\textbf{0.565}\\0.694\\2.722} & - & \tabincell{c}{0.337\\0.433\\2.481} & - \\
		%\hline
		%\tabincell{c}{Enter-\\tainment} & \tabincell{c}{0.468\\0.612\\2.590} & \tabincell{c}{0.579\\0.686\\2.548} & - & - & \tabincell{c}{0.355\\0.568\\2.600} & - \\
		\hline			
	\end{tabular}
	\vspace{-5pt}
\end{table}

\section{Acknowledgment}

This work was supported in part by the following U.S. NSF grants:  BIGDATA 1838145 and  1838147.

\section{Appendix}
\subsection{Rate Models Built on Different Values of $\alpha$}
To investigate the characteristics of rate models across news outlets, we build rate models with the observed \texttt{rate}'s among the first $\alpha$ ($\alpha$ = 5, 10, 15, 20, and 50) user comments, for both the global and local settings with Random Forest. The slopes and intercepts of regression lines based on different values of $\alpha$ are provided in Table \ref{tab:slopesInterceptsForI}.

\begin{table*}[h]
	%\small
	\centering
	\caption{Slopes and Intercepts of regression lines for rate models based on different values of $\alpha$.}
	\label{tab:slopesInterceptsForI}
	\vspace{-2mm}
	\begin{tabular}{|c|c|c|c|c|c|c|c|c|c|c|}
		\hline
		\multirow{2}{*}{} & \multicolumn{2}{c|}{$\alpha$=5} & \multicolumn{2}{c|}{$\alpha$=10} & \multicolumn{2}{c|}{$\alpha$=15} & \multicolumn{2}{c|}{$\alpha$=20} & \multicolumn{2}{c|}{$\alpha$=50} \\\cline{2-11}
		& Slope & Intercept & Slope & Intercept & Slope & Intercept & Slope & Intercept & Slope & Intercept \\
		\hline
	    Washington Post	&  0.788 & 2.692 & 0.758  &2.740	&  0.728 &2.741	& 0.700  &2.735	&  0.574 & 2.725 \\
		\hline
		Daily Mail	&  0.702 &2.553	&  0.703 & 2.606	&  0.707 &2.595	&  0.684 &2.589	&  0.595 & 2.575 \\
		\hline
		Wall Street Journal	&  0.869 &2.812	&  0.841 & 2.885 &  0.809 & 2.886 &  0.779 & 2.876	&  0.665 & 2.856 \\
		\hline
		Fox News & 0.993  &	3.217 &  0.963 & 3.201	&  0.915 & 3.149	&  0.893 & 3.113	&  0.736 & 2.973 \\
		\hline
		the Guardian	&  0.609 & 2.722	& 0.656  & 2.728	&  0.657 & 2.713	& 0.651   &2.693	&  0.566 & 2.668 \\
		\hline
		New York Times	&  0.699 & 2.933	&  0.707  & 2.935	& 0.701  & 2.910	& 0.684   &2.895	&  0.603  & 2.815 \\
		\hline
		Overall	& 0.805 & 2.744 & 0.777  & 2.767 & 0.739  & 2.749 &  0.713 & 2.738	&  0.594 & 2.698 \\
		\hline
	\end{tabular}
\end{table*}

According to the results in Table \ref{tab:slopesInterceptsForI}, the behavior observed for $\alpha = 10$ is observed for the rest of the values of $\alpha$. We summarize the key observations below:
\begin{itemize}%[noitemsep, topsep=0pt]
    \item The slope and intercept of the regression line at Fox News are the largest.
	\item The slope of the regression line at the Guardian is the smallest.
	\item Regression lines at Daily Mail and New York Times are almost parallel, the intercepts at New York Times are larger than those at Daily Mail.
\end{itemize}

Therefore, our findings in this paper for $\alpha = 10$ hold for any $\alpha \in [5, 10, 15, 20, 50]$.
%we draw the same conclusions from the analysis of rate models across outlets no matter what value of $\alpha$ is used to build the rate models.

\bibliography{rate}
\bibliographystyle{aaai}

\end{document}